\documentclass[reprint, prx, aps, notitlepage, nofootinbib,longbibliography,floatfix, superscriptaddress]{revtex4-1}
\usepackage{amsmath,amstext,amssymb,graphicx,bm,dcolumn,color,times, braket,caption}
\usepackage[colorlinks=true,citecolor=blue,urlcolor=red]{hyperref}
\usepackage[table]{xcolor}
\usepackage{graphicx}
\usepackage{environ}
\usepackage{amssymb}
\usepackage{amsmath}
\usepackage{amsthm}
\usepackage{amsfonts}
\usepackage{braket}
\usepackage{lineno}
\usepackage{subfig}
\usepackage{float}

\NewEnviron{resizeenv}{\resizebox{\linewidth}{!}{\BODY}}
\usepackage[normalem]{ulem}

\begin{document}

\title{A quantum-assisted master clock in the sky:  global synchronization from satellites at sub-nanosecond precision}

\author{Sage Ducoing}\email{sducoi1@lsu.edu}\affiliation{Department of Physics and Astronomy, Louisiana State University, Baton Rouge, LA 70803, USA}\author{Ivan Agullo}\affiliation{Department of Physics and Astronomy, Louisiana State University, Baton Rouge, LA 70803, USA}
\author{James E. Troupe}\affiliation{Xairos Systems, Inc., Denver, CO 80124, USA}
\author{Stav Haldar}\email{hstav1@lsu.edu}\affiliation{Department of Physics and Astronomy, Louisiana State University, Baton Rouge, LA 70803, USA}

\begin{abstract}
This article develops a protocol to synchronize clocks on board a network of satellites equipped with quantum resources.  We show that, in such a constellation, satellites reinforce each other's sync capabilities, forming a common clock that is  more stable and precise than its constituents. We envision the resulting network as a master clock able to distribute time across the globe,  providing the basis for a future quantum global navigation satellite system or a space-based quantum network. As an example of its capabilities, we show that a constellation of 50 satellites equipped with modest quantum resources, and distributed amongst 5 orbits at an altitude of 500 km, allows synchronization of clocks spread across the globe at sub-nanosecond precision. 
\end{abstract}

\maketitle

\section{Introduction}
\label{sec:Intro}
The limits of science and technology depend to a large extent on our ability to measure, hold and distribute time at high precision. 
Space-based platforms equipped with quantum technologies have been envisaged as promising candidates to revolutionize the scale, precision and security of sync networks \cite{LSU_satsim,Sidu2015,cubesat:2020}. 

One concrete protocol for time distribution from satellites using quantum resources was recently introduced in Ref.~\cite{paper1}, based on previous work \cite{ho:09, AntiaTroupeQCS, Lee2019}. In this protocol, a satellite acts as an intermediary to synchronize  clocks located at two ground stations on Earth's surface. This protocol utilizes the interchange of pairs of entangled photons, and takes advantage of the tight time-of-birth correlations ($\lesssim 10$ femtoseconds) associated with photon pairs produced via spontaneous parametric down conversion. We refer to this as a Quantum Clock Synchronization (QCS) protocol. This QCS protocol between stationary clocks was experimentally demonstrated in Ref. \cite{Lee2019} first and more recently in Ref. \cite{spiess_exp}.

This article proposes the use of a network of satellites interconnected and synchronized by means of an extension of this protocol, as a promising infrastructure to distribute time globally across the Earth. We support the proposal on an exhaustive study of the capabilities of such a network, based on computer simulations able to recreate the dynamics of the network in real time. The analysis shows that such a constellation of inter-communicated  satellites acquires collective advantages that enhanced the ability of individual satellites to distribute time to clientele at the Earth surface. The advantages concern both 
precision and area coverage.

The  utilization of satellites as  intermediaries benefits from the 
low loss  communication channel provided by free-space \cite{LSU_satsim, Sidu2015, Oi2017,cubesat:2020,Nano_sat, Pan_exp_crypto,Pan_satellite_QCS}. This contrasts with the large losses of  traditional fiber-optic methods, which make them inefficient for communication beyond a few 100 kms \cite{Sidu2015, LSU_satsim}.

But, shifting to a satellite comes with its own difficulties. On the one hand,  satellites are in relative motion among each other and with  ground stations, and the relative velocities affect the achievable precision of the protocol. On the other hand, it is impractical to put highly stable, large atomic clocks  onboard satellites, limiting the sync capabilities of each individual satellite. Our study shows that the collective behaviour of multiple satellites brings in advantages which compensate these limitations.  

We conclude that any city on the Earth's surface can  sync continuously  throughout the day with the ``master clock'' that the  network of satellites defines,  establishing  synchronization among  ground station at the sub-nanosecond precision. Such a  precision  at  global scales is currently beyond the capabilities of  classical protocols, including the GPS, which provides  $ \gtrsim 40$ ns of precision (95\% of the time) \cite{RTI, Zhu_GPS_relativity}. Furthermore, the interchange of entangled photons can be used to provide  an added layer of security through Bell's inequality violating polarization correlations \cite{AntiaTroupeQCS,ho:09}. 

Our quantitative analysis 
is done assuming  off-the-shelf low size, weight and power (SWaP) clocks onboard satellites, aided by already available, or near-term quantum devices, such as single photon detectors and entanglement sources (spontaneous parametric down conversion, SPDC, sources). On the other hand, we introduce some approximations and simplifications in the analysis. Although the loss model, described in Section \ref{sec:3}, provides a realistic approximation for our purposes, it ignores the effects of cloud coverage, variable noise levels during the day and night, pointing errors, and atmospheric turbulence. Additionally, we approximate the satellite orbits as circular. Nonetheless, our simulation techniques are flexible enough to include these effects in an extended analyses, which we leave for future work. We do not include relativistic corrections or  corrections arising from frequency drifts between clocks arising from other sources. 
We justify these simplifications by noticing that the optimal duration of synchronization  event turns of  to be of the order of $\lesssim 1 \rm ms$,  an interval that is short enough for the accumulated effect from first order  relativistic corrections and other frequency drifts to be well below the sync precision.

For a broader perspective on clock synchronization and its role in technological and scientific applications we point the reader to Ref. \cite{white_paper} by some of us. A brief review of the precision levels and network scales achievable by other state-of-the art clock synchronization techniques \cite{opticalclocks2018, white_rabbit, rf_picosecond, Newbury2016_freq_combs, Newbury2019_quadcopter} is also provided in Refs. \cite{white_paper, paper1, paper2}. We also point the readers to other proposals for QCS, i.e., the use of quantum resources to synchronize clocks in Refs. \cite{PhysRevA.65.052317, giovannetti:01, giovannetti:04, valencia:04, Jozsa:2010, IloOkeke2020, Lanning2022}. Proposals of quantum network of clocks made in Refs.~\cite{QClock_network_sat} and \cite{Nichol2022} are particularly relevant.

The rest of this paper is organized as follows. Section \ref{sec:QCS_protocol} summarizes the basics of the concrete QCS protocol we consider and its application to moving clocks  described in \cite{AntiaTroupeQCS,ho:09,paper1,paper2}. These tools form the basis on which the rest of this article will rest. In Section \ref{sec:3} we briefly review the simulation methods we use.  In Section \ref{sec:4} we discuss the simplest case of a network with inter-satellite sync capabilities, namely a ring of synchronized satellites,  and we analyze its advantages and quantify its capabilities through relevant figures of merit. This simple configuration will form the backbone of our proposed master clock. In Section \ref{sec:5}, we increase the complexity of the network by adding satellites in different orbits, and re-evaluate figures of merit to quantify the advantage of this new functionality. Section \ref{sec:6} is devoted to provide a test case for synchronizing a global network of ground stations, by choosing six cities across the globe. Section \ref{sec:7} contains a summary and a discussion of our results, identifying some key benefits of adding inter-satellite QCS, both at the level of the single orbit and between orbits (master clock).

\section{The QCS protocol and the concept of satellite precision shadow}
\label{sec:QCS_protocol}

\subsection{The protocol}

We start by briefly describing the QCS protocol for two stationary clocks, as first introduced in \cite{ho:09}, and show how their offset can be determined using photon time-stamp cross-correlation functions. We then review how to extend the protocol to  moving clocks, and how to evaluate sync precision and  network scale.

We consider two clocks, Alice's (A) and Bob's (B), which we initially assume tick at the  the same frequency. The task of the protocol is to evaluate the constant time offset between A and B. We must thus assume the clocks to be stable enough to maintain their frequencies up to a required precision while the protocol is executed. Suppose the constant offset between A and B is equal to $\delta$, with B ahead of A. 
Both Alice and Bob are equipped with SPDC sources that generate pairs of entangled photons, and two single photon detectors to detect photons generated locally and by the other party, respectively. $t_i^{(A)}$ and  $t_j^{(B)}$ denote times at which photons detections events $i$ and $j$ occur locally at Alice's and Bob's clocks respectively. Let $\Delta t_{AB}$ and $\Delta t_{BA}$ be the travel time for light between A and B, and vice-versa, respectively. The round trip time is therefore $\Delta t=\Delta t_{AB}+\Delta t_{BA}$. The protocol described here rests on the assumption of {\em reciprocity}, namely the equality between  $\Delta t_{AB}$ and $\Delta t_{BA}$. In classical protocols using radio frequency signals, and for the case where Alice and Bob are a ground station and a satellite in a LEO (low Earth orbit), respectively, the fluctuations in signal multi-path propagation can create channel non-reciprocity of the order a nanosecond. In contrast, for optical photons this figure goes down to the order of 10's of femtoseconds \cite{Belmonte2017,Taylor2020}, which justifies our assumption.
Now, to obtain the absolute time difference between clocks, $\delta$,  consider first a photon pair produced at Alice's location. One of the photons of the  pair is detected locally at  Alice's detector and the other member of the pair travels to Bob, accumulating a travel time $\Delta t_{AB}$. For any particular pair event produced at Alice's site, the difference between the time labels recorded by Alice and Bob will be $t^{(A)}_i-t^{(B)}_i=\Delta t_{AB}+\delta$. Similarly, for any pair produced at Bob's site $t^{(B)}_j-t^{(A)}_j=\Delta t_{BA}-\delta$. The value  of $\delta$ can be obtained from the knowledge of these two time differences. In practice, however, one finds the added difficulty that Alice and Bob  receive many photons from ambient noise, and need a way to identify which photons belong to an entangled pair (several photon pairs lose one partner, therefore, also act like ambient noise). Correlation functions are a useful tool to disentangle the signal from the noise. Consider, first,  pairs produced at Alice's location. The detection times can be translated into distributions by
\begin{eqnarray}
a(t)=\sum_i \delta\left (t-t^{(A)}_i\right)  \, , \ \ \ \ 
b(t)=\sum_j \delta\left (t-t^{(B)}_j\right), 
\end{eqnarray}
where $\delta(x)$ is  Dirac's delta, and $i$ and $j$ index arbitrary detection events, which can arise either from photons belonging  to entangled pairs  or from other detector triggers such as stray light, dark counts, etc.
We now compute cross-correlation as: 
\begin{equation}\label{cAB}
c_{AB}(T)=(a\star b)(T)=\int_0^{t_{\rm Aqc}} a(t)\, b(t+T)\, \mathrm{d}t\, ,
\end{equation}
where $t_{\rm Aqc}$ is the duration of the protocol. 
The function $c_{AB}(T)$ counts the total number of detection events at Alice and Bob that are separated  precisely by an interval $T$. For a sufficiently large number of detected entangled pairs (large signal-to-noise), this correlation function would show a peak  at $T=T_{AB}=\Delta t_{AB}+\delta \rm{.}$
Likewise, if we consider  pairs created on Bob's site, we can construct another cross-correlation
\begin{equation}\label{cBA}
c_{BA}(T)=(b\star a)(T)=\int_0^{t_{\rm Aqc}} b(t)\, a(t+T)\, \mathrm{d}t,
\end{equation}
which will show a peak at $T=T_{BA}=\Delta t_{BA}-\delta \rm{.}$
From these, we can extract both the round trip time $\Delta t$ and the absolute time difference between clocks {\em without any knowledge of the length of the path between Alice and Bob}:
\begin{eqnarray}
\Delta t = T_{AB}+T_{BA}\, , \ \ \ \ \ \ 
\delta = \frac{1}{2}\left (T_{AB}-T_{BA}\right ), 
\label{eqn:offset}
\end{eqnarray}
where in the last equation we have used  channel reciprocity, $\Delta t_{AB}=\Delta t_{BA}$.

The quantum entanglement between in photon pairs created at Alice's and/or Bob's locations does not play an important role in determining $\delta$, except for the fact that the production of entangled pairs by SPDC ensures that the photons in each pair are generated within a time window typically of a few  $100$ fs---several orders of magnitude smaller than the synchronization precision we aim to obtain \cite{Shih2004}. In contrast, the entanglement between photon pairs plays an important role in increasing the security of this protocol to target malicious attacks by utilising well established quantum cryptography techniques, such as Bell tests.
\\

Additional difficulties appear when Alice and Bob are in relative motion. If the distance between the two parties change in time, detection events corresponding to different pairs of entangled photons will not accumulate in a single peak of the correlation function. The peak will spread. If the spread happens fast enough, the peak will be buried under the noise. On the contrary, if  synchronization  can be achieved before noisy events accumulate in the correlation function, the protocol would be successful. There is, therefore, a competition between the duration of the protocol,  
$t_{Acq}$, and the noise level. Note that $t_{Acq}$ cannot be made arbitrarily small, since enough photons must be collected to get statistically significant peaks for the cross-correlation functions. This effect is very similar to adding a relative frequency difference between the clocks at A and B, which has been studied in detail in Ref. \cite{ho:09}. 

 The optimal precision this protocol can achieve when applied to moving clocks can be obtained in terms of the following parameters:  the  ebit rate of the SPDC sources,  $R$; the total loss in the channel between the clocks A and B (including detector losses), $\eta$; the radial relative velocity, $v_{rel}^{rad}$; and the minimum (average) number of photons exchanged between A and B needed for the cross-correlation peaks to be statistically significant, $N_{\rm min}$. 

Denoting the optimal sync precision by  $t_{\rm bin}$, it is given by \cite{paper2}
\begin{equation}
    t_{\rm bin} = \frac{N_{\rm min}\, v_{rel}^{rad}}{c\, R\, \eta},
    \label{eqn:max_precision}
\end{equation}
where $c$ is the speed of light in vacuum. Since $t_{\rm bin}$ gives the smallest detectable time offset \cite{paper2}, 
a larger $t_{\rm bin}$ means a worse sync precision. 

Note that link transmissivity $\eta$ and the relative radial velocity $ v_{rel}^{rad}$ are  functions of the distance between A and B and, consequently, they change continuously in time. Throughout this paper,  we will use Eqn.~\eqref{eqn:max_precision} to determine the precision at which two moving clocks can synchronize, applied to either satellite-satellite or satellite-ground station. Distances between the parties will be  computed by numerically simulating the dynamics of  a constellation of satellites distributed in different orbits. 

\subsection{Precision shadow and sync traces}

We introduce next a visual tool that becomes handy in determining   the limits of performance and scale of the QCS network. We use the term ``satellite precision shadow'' for this tool. Given a satellite, the level of precision at which it can sync with ground stations in Earth's surface is determined by the distance between the two parties, as we just explained above. Therefore, if we fix a value of the desired  sync precision, $t_{\rm bin}$, Eqn.~\eqref{eqn:max_precision} can be used to determine the patch on Earth's surface under the satellite at which synchronization at such precision can be established. This is patch defines what we call   {\em precision shadow} of the satellite. Any two ground stations located within the precision shadow at an instant can sync at optimal precision $t_{\rm bin}$ using the satellite as a mediator.

But the previous definition of precision shadow does not take into account the precision of the clock on board the satellite. Improving the precision of that clock implies that a satellite can ``hold time'' for longer and, consequently, be able to sync cities that are further apart. In other words, the size of the precision shadow increases, elongating along the direction of the motion of the satellite. The precision of the satellite's clock can be quantified by means of the 
``holdover time'' $\tau$, defined as the average interval it takes for the clock to accumulate a deviation equal to the desired sync precision $t_{\rm bin}$. 

\begin{figure}[ht]
\centering
    \includegraphics[width=0.8\linewidth]{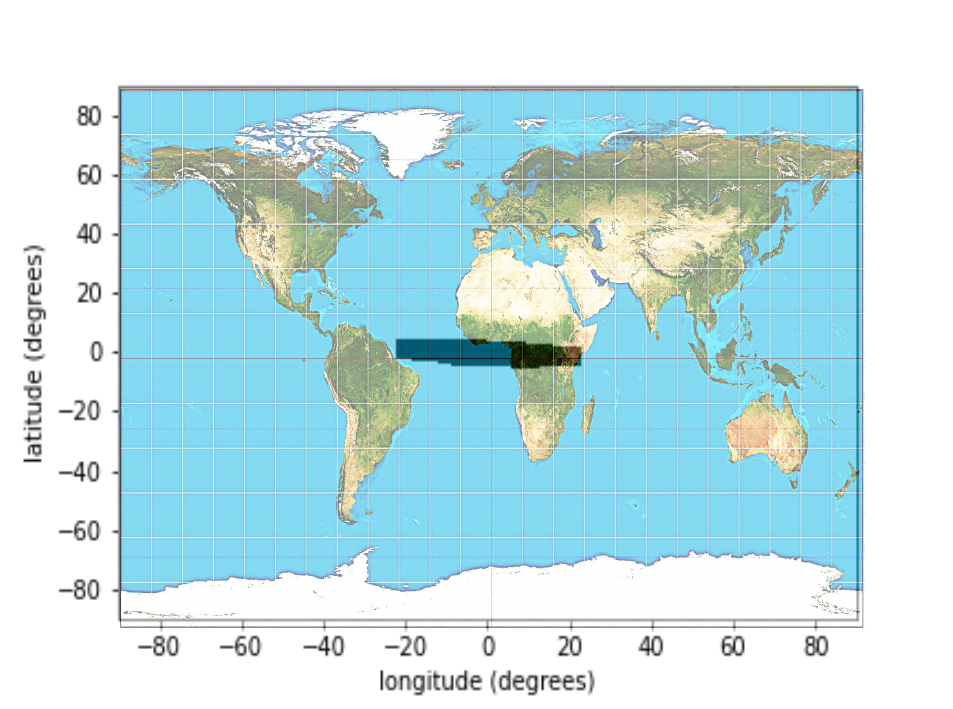}
    \includegraphics[width=0.8\linewidth]{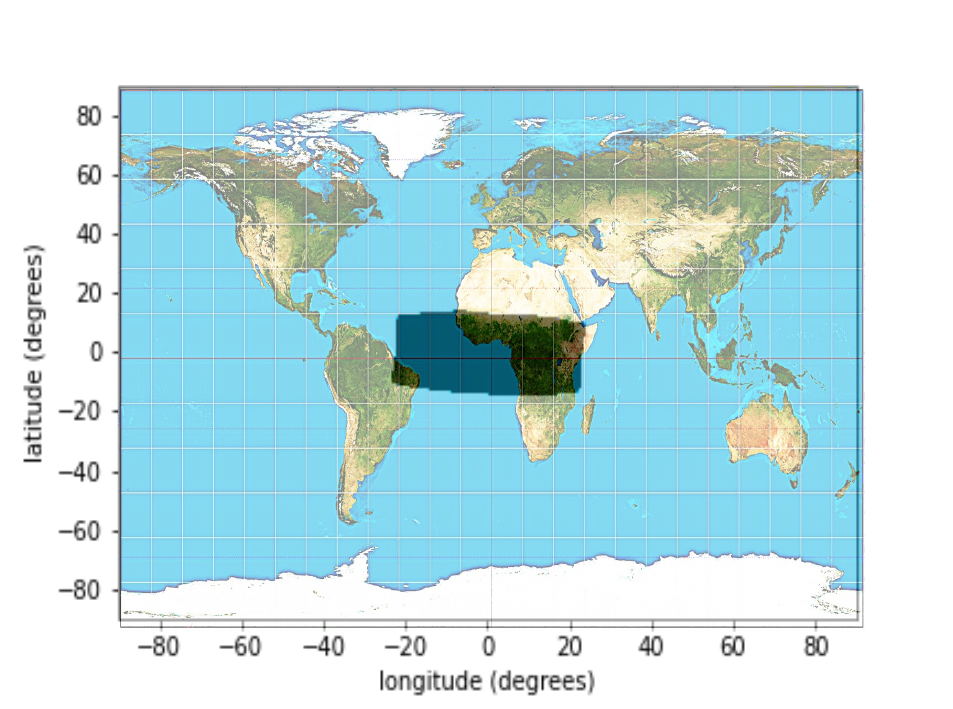}
    \includegraphics[width=0.8\linewidth]{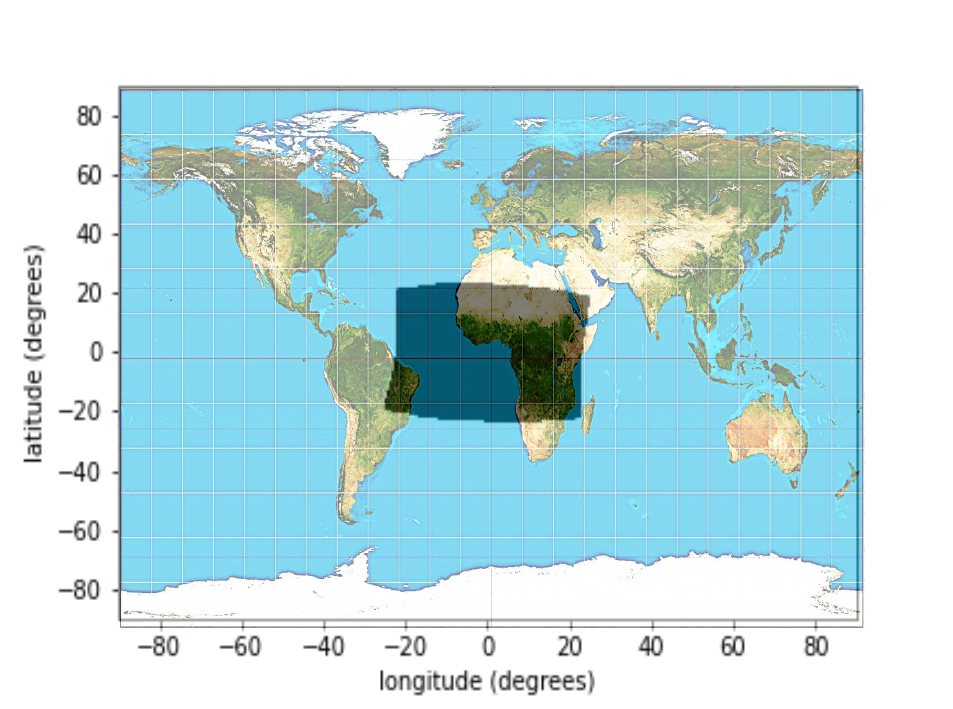}
    \caption{1 ns precision shadow of a single LEO satellite in a 500 km altitude polar orbit when it is located just above the intersection of the Prime Meridian and the Equator. All points within the grey shadow can sync with the satellite at a precision better than 1 ns at this instant of time. As the holdover increases from $\tau = 0$ min (top) to $\tau = 5$ min (center) and finally $\tau = 10$ min (bottom), the shadow elongates along the satellite's direction of motion, increasing the network scale.}
    \label{fig:single_sat_shadow}
\end{figure}

As an example, Fig.~\ref{fig:single_sat_shadow} shows the 
 precision shadow of a satellite in a 500 km LEO polar orbit, when the satellite is just above the intersection of the Prime Meridian and the Equator. We have chosen $t_{\rm bin}=1\, ns$ as sync precision target. The plot shows the effect of increasing the satellite's clock  holdover times $\tau$. Details of the simulations are spelled out in more detail in the next section.

\begin{figure}[ht]
    \centering
    \includegraphics[width = 0.9\linewidth]{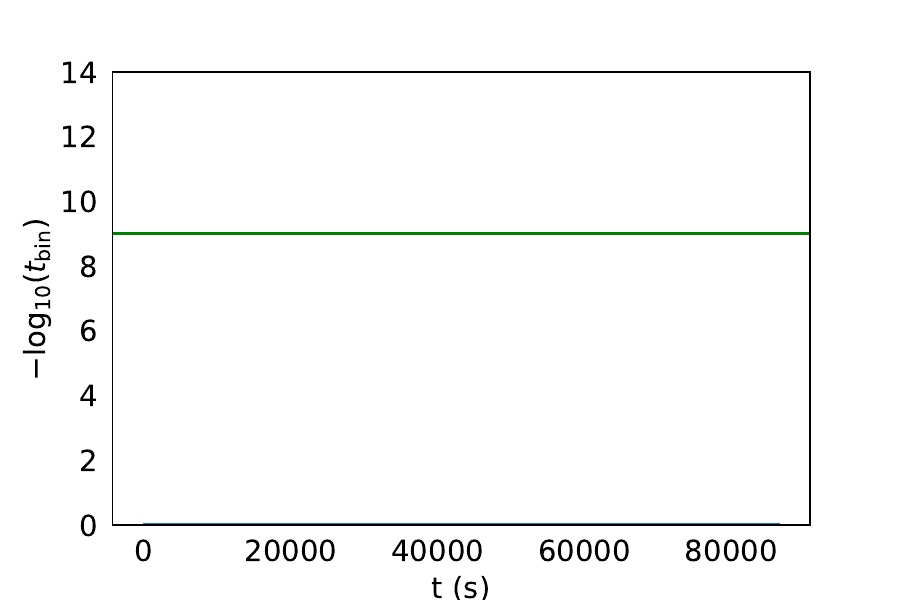}
    \includegraphics[width = 0.9\linewidth]{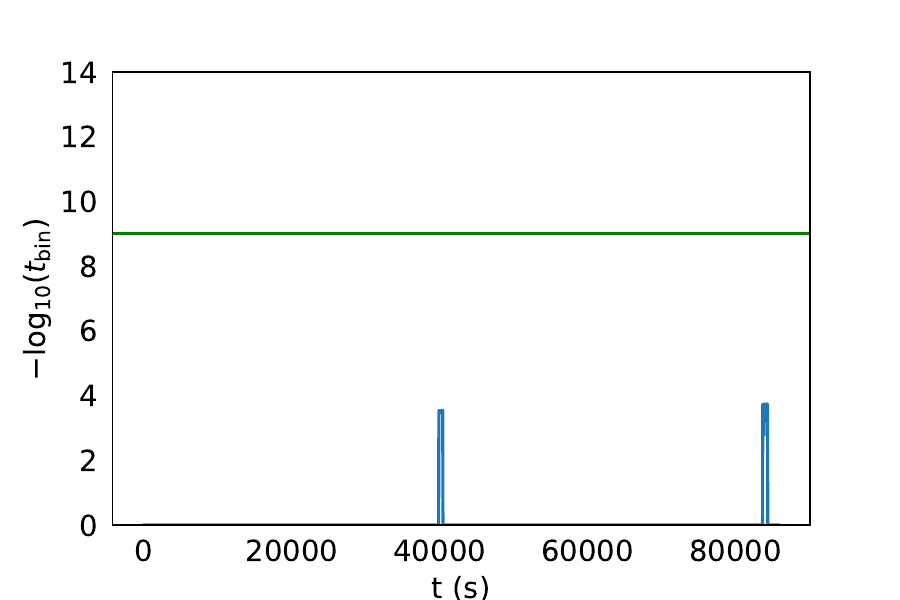}
    \includegraphics[width = 0.9\linewidth]{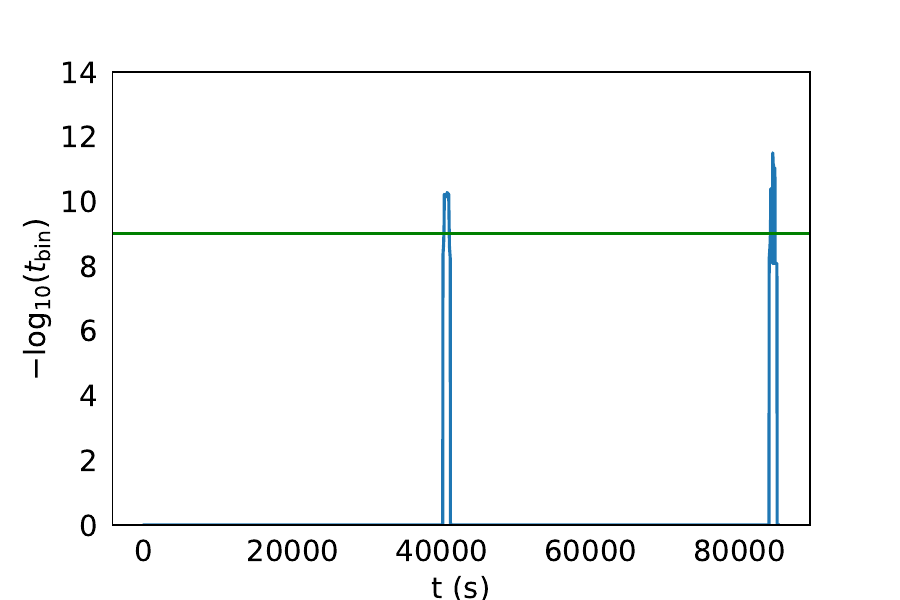}
    
    \caption{Sync traces between New York and Salt Lake City, using a single LEO satellite in a 500 km polar orbit, both in terms of precision and the number of connections every day. As the holdover time $\tau$ is increased from (top) $\tau = 0$ min to (center) $\tau = 5$ min, and finally to (bottom) $\tau = 10$ min, we start seeing sync events, first at worse than nanosecond precision, but then, for 10 min holdover, we see sub-nanosecond level sync precision. The green (horizontal) line indicates the cut-off introduced by the satellite clock's precision within the holdover time, which has been chosen to be 1 ns.}
    \label{fig:holdover_NY_SLC}
\end{figure}

To illustrate the application of the sync protocol under consideration via our numerical simulations, we choose two cities, New York and Salt Lake City in this example, and compute  the maximum precision at which they can sync by means of a single LEO  satellite in a 500 km
polar orbit, as a function of time. We call this type of  plot \emph{sync trace} \cite{paper1}, since it monitors the sync precision between two ground stations in the course of time. 
Fig.~\ref{fig:holdover_NY_SLC} shows the  sync trace between New York and Salt Lake City. The maximum sync precision at a given instant $t$ is determined as the minimum value of the two precisions at which the ground stations can individually sync with the satellite, within an interval $\tau$ around $t$ (recall, $\tau$ is the satellite's clock holdover). Obviously, the maximum sync precision achievable is given by the precision of the satellite's clock---assuming that the ground stations use better clocks. This absolute maximum precision is indicated by a horizontal green line in   Fig.~\ref{fig:holdover_NY_SLC}, which we take to be 1 ns in this example. This maximum  can be improved just equipping  satellites with better clocks. As an aid to intuition, we mention here that off-the-shelf Rubidium atomic clocks can hold time to 1 ns precision for roughly $10$ minutes \cite{DARPA_ppt_2019}. This will be a common feature of all sync traces shown in this paper.\footnote{The precision of the clock on board satellite can be quantified by the Allan deviation curve \cite{oelker:04}, which is the standard measure of a clock's stability is often obtained empirically. Here, for our simulations it suffices to take a simplistic, worst case scenario approach, corresponding to precision $C$  for an interval $\tau$, and zero beyond that interval.}

The improvement in sync outcomes as the holdover time $\tau$ is increased for the New York-Salt Lake City ground station pair ($\approx$ 3500 km apart) can be seen in Fig.~\ref{fig:holdover_NY_SLC}. For $\tau = 0$ there was no connection due to visibility constraints. As holdover increases, and thus also the extent of the satellite shadow, some low-precision sync is possible twice per day. For $\tau=10$ minutes we finally see the capability of sub-nanosecond sync between the two cities. 

Having established the capabilities of a single satellite, we will now investigate the effects of adding satellites, both with and without the ability to synchronize amongst themselves, on ground-level sync outcomes.

\section{Simulation methods}
\label{sec:3}

Our simulation consists mainly of two parts: (1) simulating the dynamics of the satellite constellation and ground the stations and; (2) simulating the lossy quantum communication channel between the two parties attempting synchronization.

Regarding  dynamics, we simulate the trajectory of a constellation of satellites following circular orbits (extension to non-circular is possible), and compute at every time step of the simulation the relative positions and distances of each satellite and ground station. As described below, we use these distances to compute the efficiency of the  free space communication channel. 

Furthermore, we check the visibility of satellites from ground stations and satellites among themselves, in the following manner. For a ground station satellite pair, we check that the angle between the line joining the ground stations to the satellite and the ground's station zenith is less than $\pi/2$.  Whereas for satellite-satellite visibility, the distance from the center of the Earth to the center of the line joining the two satellites,  must be greater than $R_E+\Delta_{\rm th}$, where $\Delta_{\rm th}$ is the atmospheric thickness. We assume $\Delta_{\rm th}= 10$ km. This condition makes the satellite-satellite links impervious to effects of atmospheric scattering.

From the position of each satellite and ground station, we  can  evaluated the relative radial velocity between any parties at every time step in the simulation.  The relative radial velocity between two parties (A,B) is given by:
\begin{equation*}
     v_{\rm rel}^{\rm rad} = (\mathbf{r_{A}} \times \mathbf{\omega_{A}} -  \mathbf{r_{B}} \times \mathbf{\omega_{B}}) . \frac{(\mathbf{r_{A}} - \mathbf{r_{B}})}{\vert (\mathbf{r_{A}} - \mathbf{r_{B}}) \vert} \, ,
\end{equation*}
where, $\mathbf{r_A}, \mathbf{r_B}$ are their position vector with respect to  the center of the Earth,  and $\mathbf{\omega_{A}}, \mathbf{\omega_{B}}$ are their angular velocities.

Regarding the communication channel, we take into account the various losses by calculating effective efficiency factors. For the ground station-satellite links, we must distinguish between the uplink (ground station to satellite) and downlink (satellite to ground station) $\eta_{(up)}$ and $\eta_{(dwn)}$ respectively, since we assume different telescope radii for the two (atmospheric turbulence and scattering effects would also be slightly asymmetric, but our model does not include these details). $\eta_{(up)}$ is the probability that a photon pair generated out of the SPDC source at the ground station will lead to a double detection event, i.e., one partner will be detected at A locally and the other at B after travelling through free space and the atmosphere. Similarly for $\eta_{(down)}$. For satellite-satellite links $\eta_{(up)} = \eta_{(down)}$.

We model the satellite-ground quantum communication channel as follows. For concreteness, let us focus on a downlink channel. The following will hold similarly for an uplink channel. We assume clear skies and approximate the downlink channel as only lossy (background noise is accounted for at the detectors). That is, photons are either transmitted through the channel or lost in transmission. The dominant sources of loss are (1) beam spreading (free-space diffraction loss), (2) atmospheric absorption/scattering, and (3) non-ideal photodetectors on the satellite and on the ground. We characterize these loss mechanisms by their transmittance values, which is the fraction of the received optical power to the transmitted power (which is also equal to the probability to transmit/detect a single photon). Let these transmittance values be, respectively,
\begin{equation*}
  \eta^{(dwn)}_{fs}(L), \quad \eta^{(dwn)}_{atm}(L), \quad \kappa_{sat}, \quad \kappa_{gs}\, ,
\end{equation*}
where the superscripts refer to the downlink, $fs$ refers to free-space diffraction loss, $atm$ to atmospheric loss, $L$ is the link distance (physical distance) between satellite and receiver (which in turn depends on the satellite altitude, $h$, position of the satellite in its orbit, and position of the ground station on Earth's surface), $\kappa$ denotes non-ideal detection efficiencies for the onboard satellite detector array ($sat$) as well as the detector array at the ground station ($gs$), and all transmittance values are less than or equal to $1$. Simple analytic formulae are used to estimate the free-space and atmospheric transmittance values in accordance with \cite{LSU_satsim}.
Therefore, the overall efficiencies are given as:
\begin{eqnarray}
\label{eqn:39-40}
    \eta_{(dwn)} &=& \eta^{(dwn)}_{fs}\eta^{(dwn)}_{atm}\kappa_{sat}\kappa_{gs}\, , \\
    \eta_{(up)} &=& \eta^{(up)}_{fs}\eta^{(up)}_{atm}\kappa_{sat}\kappa_{gs}\, . 
\end{eqnarray}
The satellite-satellite link efficiencies can also be similarly evaluated. One key difference, given our assumptions, is that these channels are free of atmospheric scattering, i.e. $\eta^{(up)}_{atm} = \eta^{(dwn)}_{atm} = 1$.

For convenience, we list in Table \ref{tab:table1} other operational parameter entering in  our simulations along this paper, and the  values for them we have used to illustrate our techniques.

With this information, we use Eqn.\eqref{eqn:max_precision} to evaluate the time sync precision between the two parties at any time step of the simulation.

\begin{table}[]
\label{tab:table1}
\centering
\begin{tabular}{|l|l|}
\hline
Altitude of orbits                              & h = 500 km         \\
\hline
Operational wavelength                            & $\lambda$ = 810 nm         \\
\hline
Radii of telescopes                  & ($r_{sat}$, $r_{gs}$) = (10 cm, 60 cm) \\
\hline
Detector efficiencies                        & ($\kappa_{sat}$,$\kappa_{gs}$) = (0.5, 0.5)     \\
\hline
Source rate                                       & $R$ = $10^7$ entangled-pairs/s   \\
\hline
\end{tabular}
\caption{Various operational parameters for the simulations}
\centering
\label{tab:table1}
\end{table}

\section{Single orbit master clock}
\label{sec:4}

\begin{figure*}[ht]
    \centering
    \includegraphics[width=0.6\linewidth]{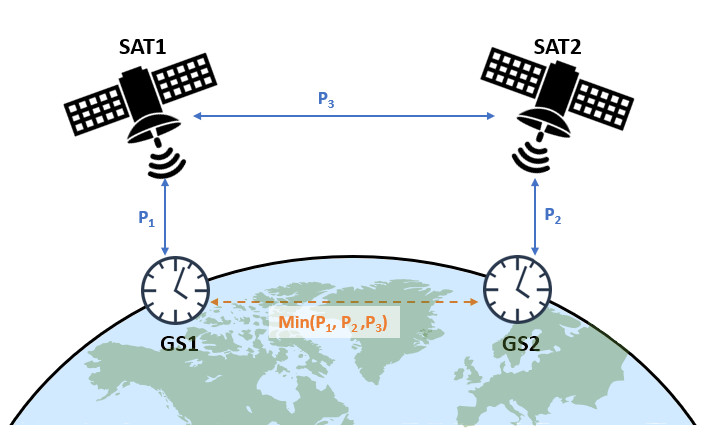}
    \caption{Schematic for a 3-link time sync involving two satellite-ground station links and one satellite-satellite link. Each link has sync precision $P_{i}=-{\rm log_{10}} (t_{bin,i})$, and the effective sync precision between the two ground station clocks (\textbf{GS1/GS2}) is restricted to the minimum precision over all links. These satellite-satellite links can arise through either intra-orbit or inter-orbit synchronization, which are explored in  detail in Sections \ref{sec:4} and \ref{sec:5}.
}
    \label{fig:3link}
\end{figure*}

In this section, we begin by discussing the simplest case of inter-satellite synchronization: synchronizing a ring of satellites following the same orbit. 
Inter-satellite synchronization  is illustrated in Fig.~\ref{fig:3link}. It will form the backbone of our proposed master clock.

\subsection{Intra-orbit synchronization}
Consider a ring of satellites in circular orbit around the Earth. When the satellites do not sync with one another, for two cities to synchronize they must  simultaneously lie within the precision shadow of the same satellite. This requirement greatly constrains the possibilities of synchronization, given the limited size of the shadows. It is challenging to sync cities separated by more than $\approx$ 1000 km, unless satellites are equipped with high quality clocks (see Figs.~\ref{fig:single_sat_shadow}). Another option is allowing the (poor/medium quality) satellite clocks to synchronize with each other. With this functionality, for two cities to synchronize, they do not need to connect to a common satellite, but must sync to  {\em any} satellite in the constellation within a time window $\tau$. 

In this subsection, we analyze the necessary conditions for adding this functionality ---intra-orbit synchronization--- as well as the consequent improvement in precision and network scale of the QCS network. 

An obvious condition for intra-orbit sync is satellite to satellite \textit{visibility}. If we place the first satellite at the point (0, 0, $R_{E}+h$), with respect to the Earth's center, and where $R_{E}$ is the Earth's radius, the angle the second satellite makes with the positive $z$-axis must be less than $\pi- 2\theta_{s}$. Here $\theta_{s}$, which we call the shadow angle, is
\begin{equation}
    \theta_{s} = \arcsin(\frac{R_{E}}{R_{E}+h}).
\end{equation}
\\
 If we consider polar orbits at $h=500$ km altitude, this corresponds to $\theta_{s} \approx 68^\circ$. For an orbit of equally spaced satellites at this altitude, nine satellites suffice for
 the visibility condition to be satisfied at any instant, and each satellite is always in view from the nearest neighbor. Increasing the orbital height decreases the required number of satellites for continuous  visibility.

For two satellites A and B in the same (circular) orbit, the orbital velocities are equal, i.e., $\mathbf{\omega_{A}} = \mathbf{\omega_{B}}$, so $v_{\rm rel}^{\rm rad}=0$. Therefore, according to Eqn.~\eqref{eqn:max_precision}, the sync precision has no lower bound and is only limited in practice by the time-stamping precision. Further, once nearest-neighbor satellites sync with each other, all satellites become synchronized transitively. Thus, to form a master clock comprised of a ring of polar satellites sharing a common time, the visibility condition becomes the only requirement which, as mentioned above, is solved by adding enough satellites to the orbit.  

\subsection{Synchronization precision and network size}
Once the visibility condition between satellites has been established, the task of syncing two ground stations GS1 and GS2 via two satellites SAT1 and SAT2 proceeds as follows (see Fig.~\ref{fig:3link} for an illustration). 

We first find the minimum precision that GS1 can get at time $t$ from all satellites in the orbit, and then the minimum precision that GS2 can get from the orbit within a period $(t-\tau/2, t+\tau/2)$, where $\tau$ is the holdover time of clocks on board satellites (assuming all identical). The larger of these two values is the sync precision for the city pair. This protocol uses that satellites can sync among themselves at better precision than satellites with ground stations, because they do not move relative to each other.

\begin{figure}[ht]
\centering
    \includegraphics[width=0.9\linewidth]{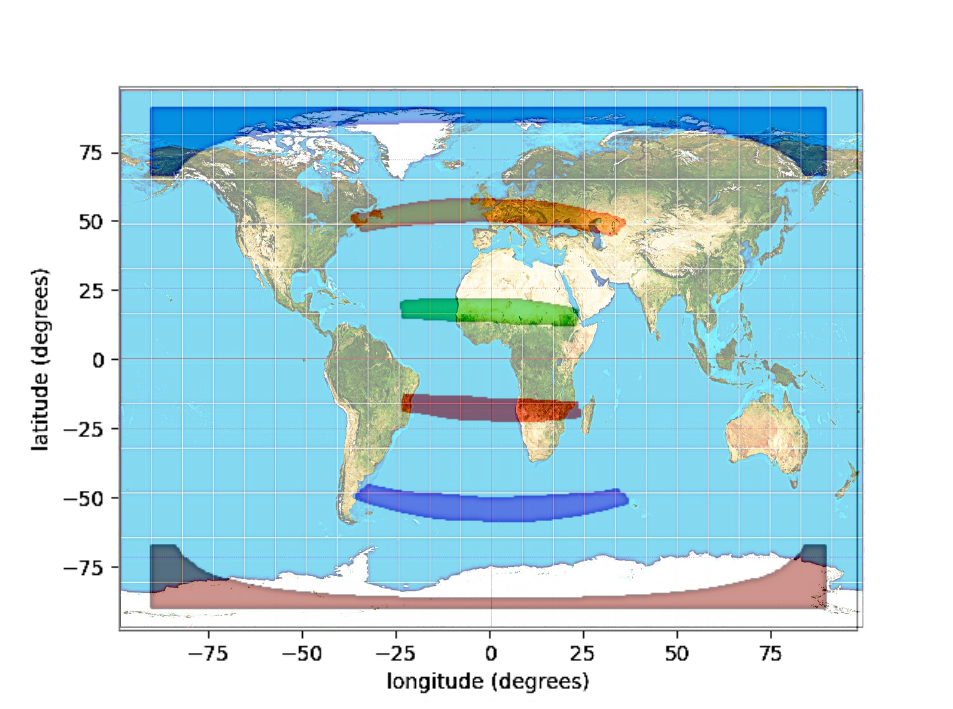}

    \includegraphics[width=0.9\linewidth]{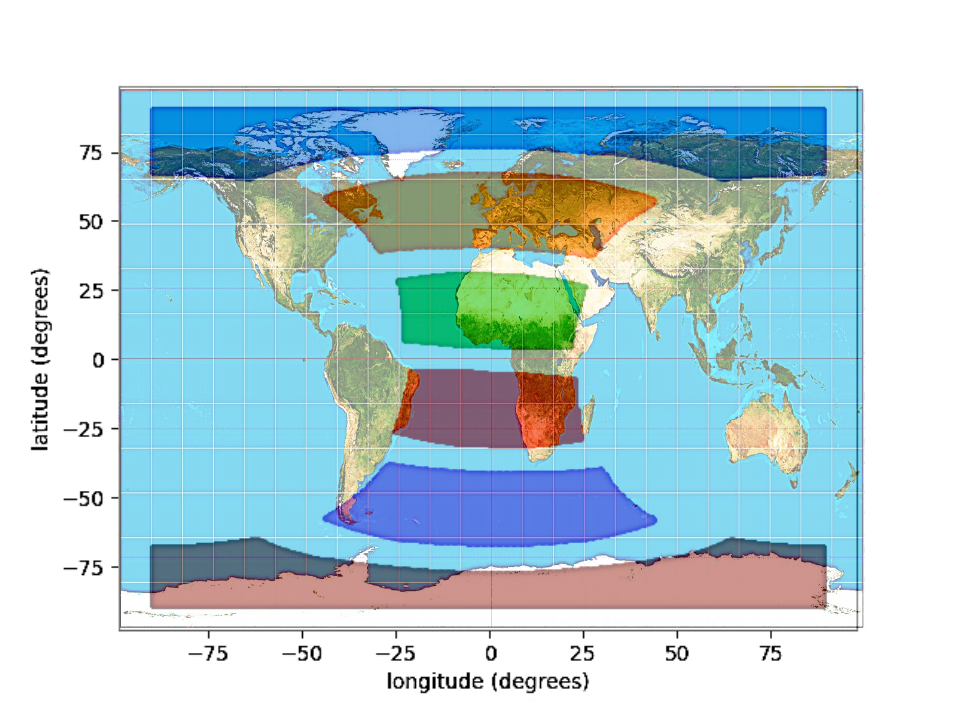}
    \includegraphics[width=0.9\linewidth]{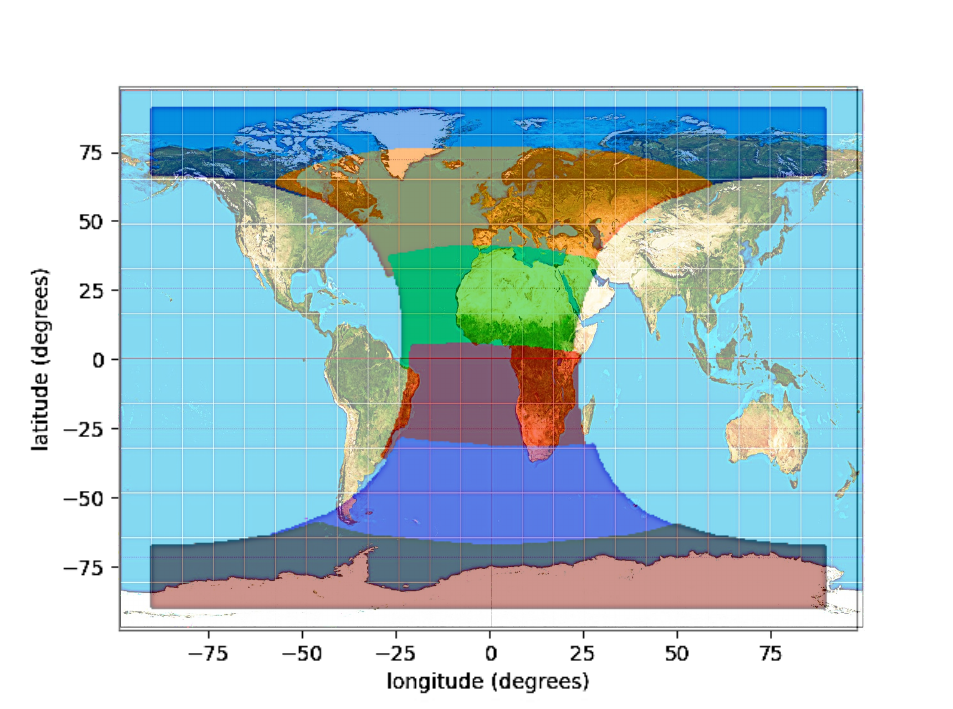}
    
    \caption{1 ns precision shadows for a constellation of 10 satellites in a 500 km polar orbit along the Prime Meridian. In this plot, the satellites are not in sync with each other, and consequently the different patches are independent. Each plot shows a different value of the  holdover times: $\tau = 0$ min (top), $\tau = 5$ min (middle), and $\tau = 10$ min (bottom).}
    \label{fig:shadow_disconn}
\end{figure}

\begin{figure}[ht]
\centering
        \includegraphics[width=0.9\linewidth]{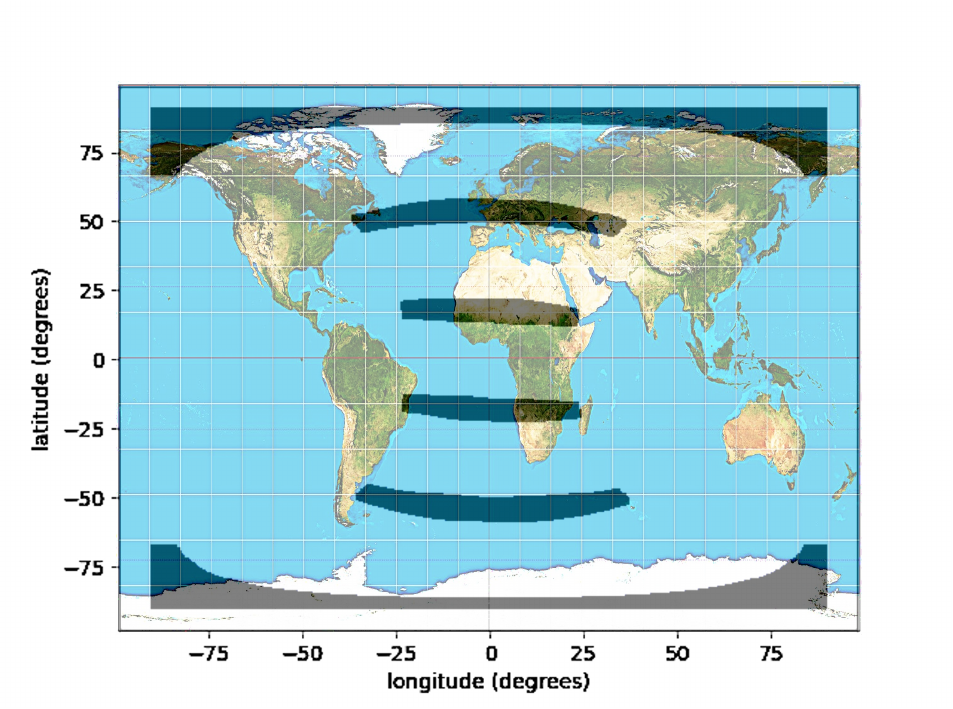}
        \includegraphics[width=0.9\linewidth]{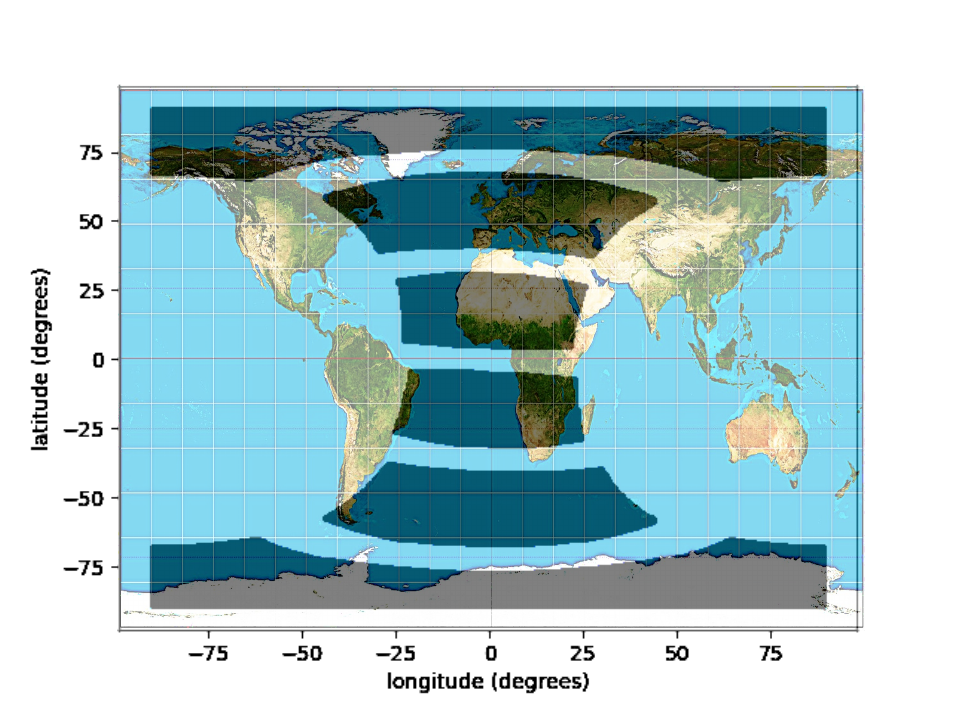}
        \includegraphics[width=0.9\linewidth]{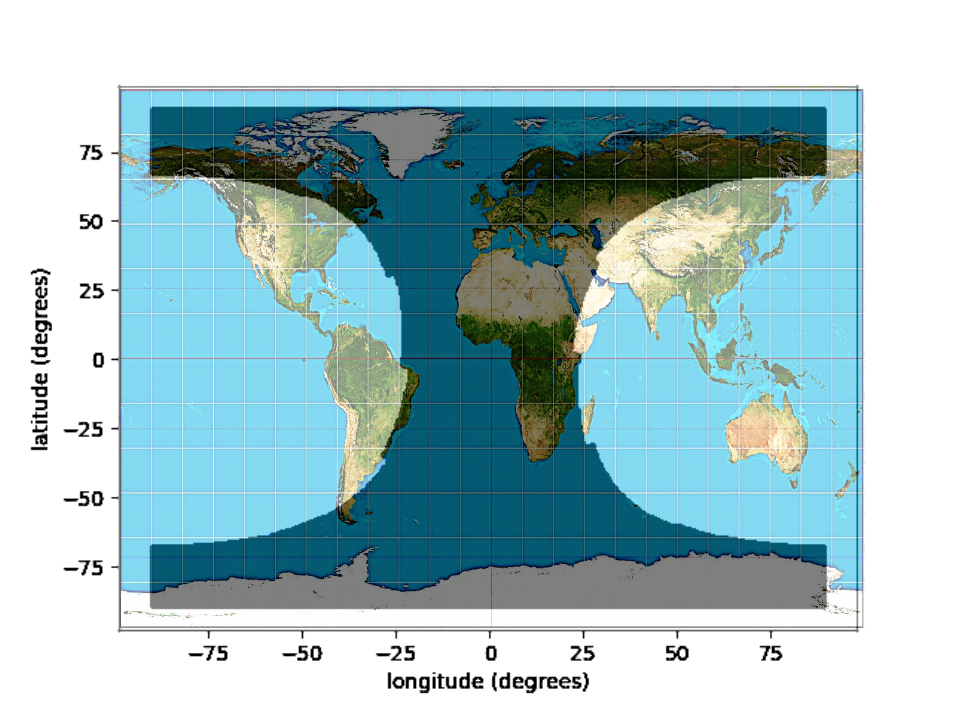}
        
    \caption{1 ns sync precision shadow for an orbit of 10 satellites in a  500 km altitude polar orbit along the Prime Meridian. The satellites are synchronized with each other, and consequently the the individual patches 
    are part of a unique shadow. Different  holdover times are shown: $\tau = 0$ min (top), $\tau = 5$ min (middle), and $\tau = 10$ min (bottom).}
    \label{fig:shadow_conn}
\end{figure}

We  consider now an example of polar orbit with 10 satellites at 500 km altitude. In Figs.~\ref{fig:shadow_disconn} and \ref{fig:shadow_conn}, we compare the regions of achievable nanosecond sync precision for a disconnected orbit --- no inter-satellite sync --- to an orbit with intra-orbit sync functionality. We show the result for different holdover times $\tau = 0, 5,$ and $10$ min. The differently colored precision shadows in Fig.~\ref{fig:shadow_disconn} indicate that they are disconnected and independent shadows,  hence sync between two cities occurs only when they fall within one single shadow patch (one single color). Once the intra-orbit synchronization functionality is added (Fig. \ref{fig:shadow_conn}), all shadow patches merge together to form a single bigger shadow. With this functionality established, it is clear that the network scale (geographic extent) is enhanced substantially. Without any holdover capabilities, each shadow is $\approx 8^\circ$ in width, and increases by about $3.5^\circ$ per added minute of holdover. We see in Fig.~\ref{fig:shadow_conn} that in order to have continuous latitudinal connection (bottom plot) with this configuration of ten satellites, we require 10 minutes of holdover time. 

Before discussing detailed examples, we further stress here the advantage achieved via intra-orbit connectivity. Equipped with modest clocks with 5-10 mins of holdover (off-the-shelf Rubidium atomic clocks) we can provide high precision synchronization between cities on the opposite ends of the globe (latitude-wise). For example, 1 ns sync precision is achievable between New York City in the US and Puerto Montt in Chile with 5 mins of holdover time (see Fig.~\ref{fig:const-long-5minprec} (middle)), and between New York City and Beijing, China, with 10 mins of holdover time (see Fig.~\ref{fig:const-latitude} (top)), with the simple 1 orbit configuration discussed above. This corresponds to a sync network range of roughly 10000 km (considering geodesic distances between the cities). In contrast, for disconnected satellites this scale could not surpass 4000 km with similar holdover times. The intra-orbit functionality changes the constellation capabilities  from continental to global scale.

\subsection{Sync Examples}
\begin{figure*}[ht]
    \centering
    
        \includegraphics[width=0.45\linewidth]{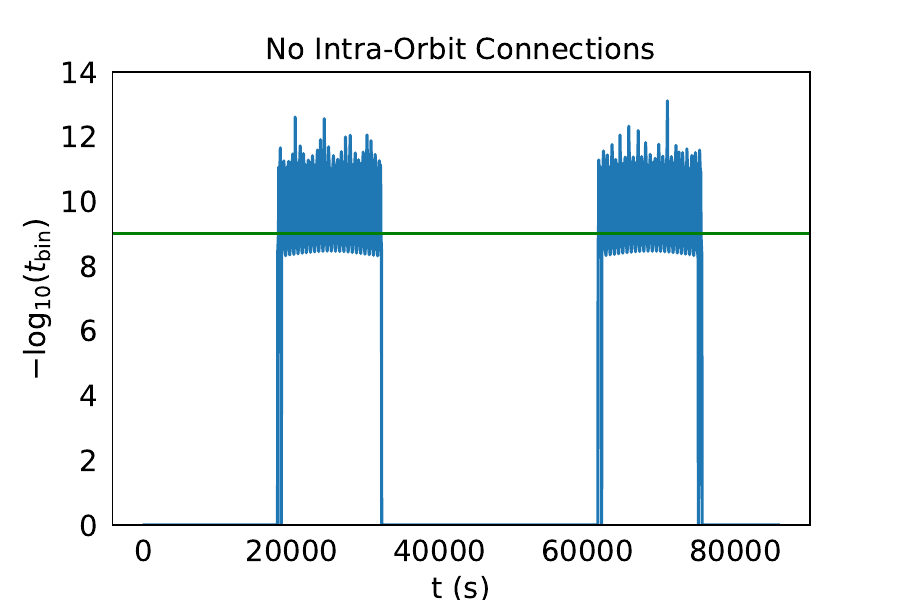}
        \includegraphics[width=0.45\linewidth]{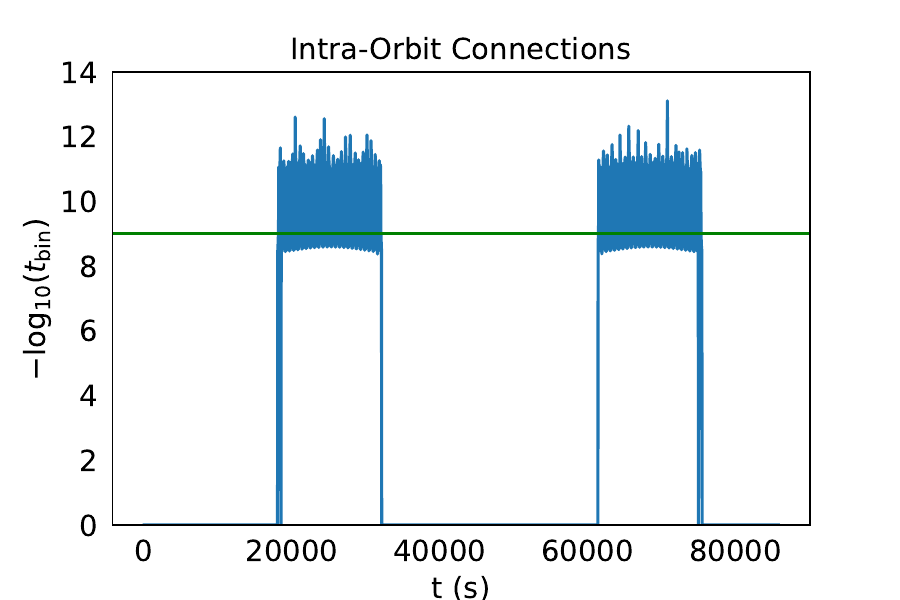}

        \includegraphics[width=0.45\linewidth]{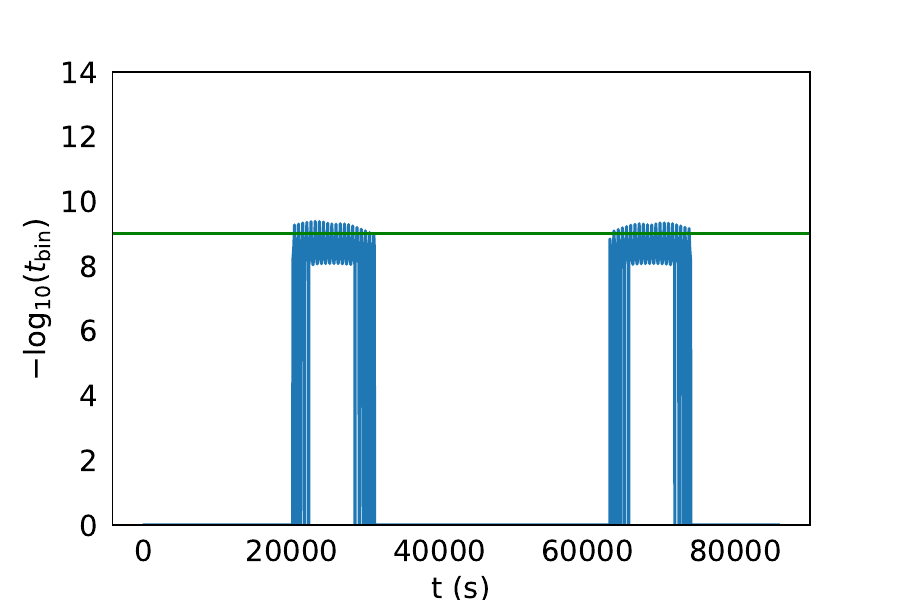}
        \includegraphics[width=0.45\linewidth]{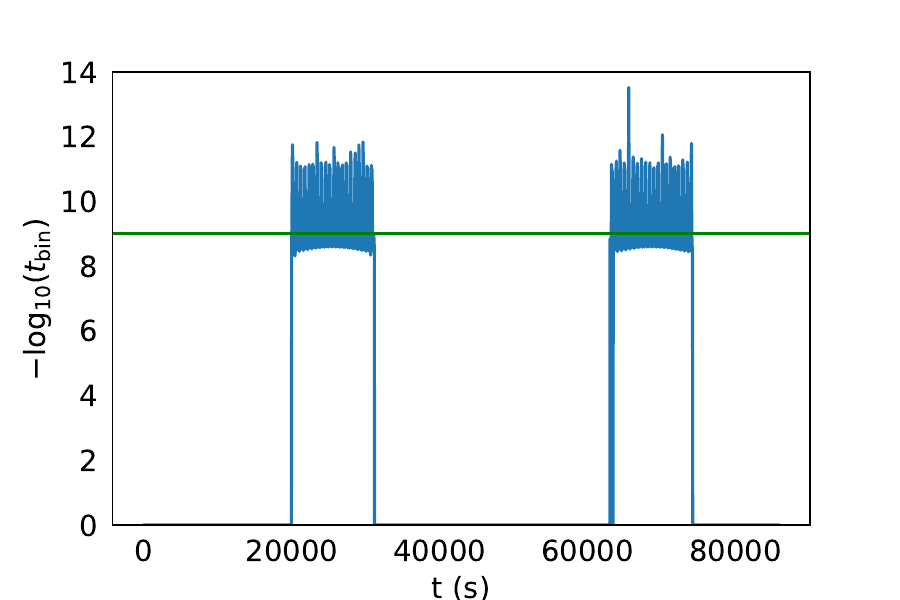}

        \includegraphics[width=0.45\linewidth]{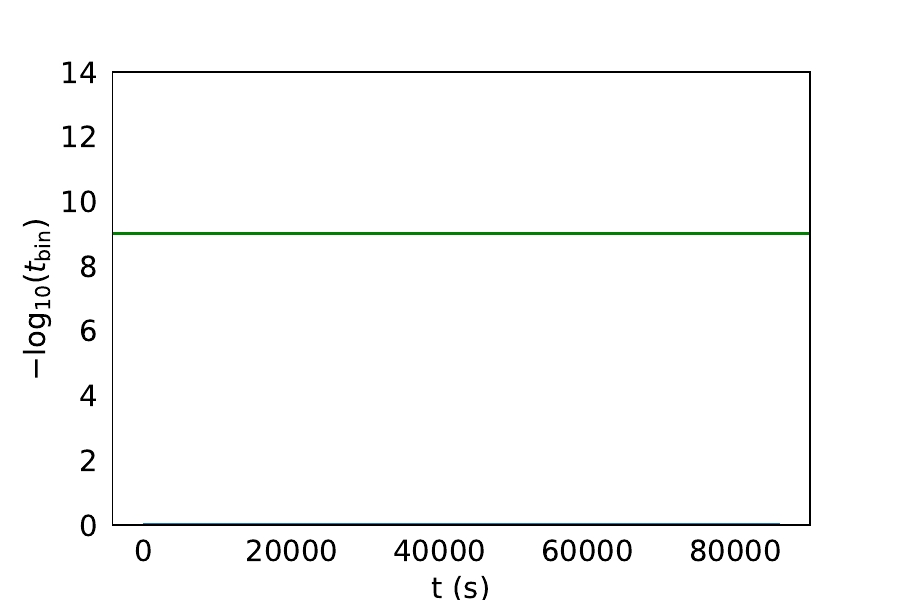}
        \includegraphics[width=0.45\linewidth]{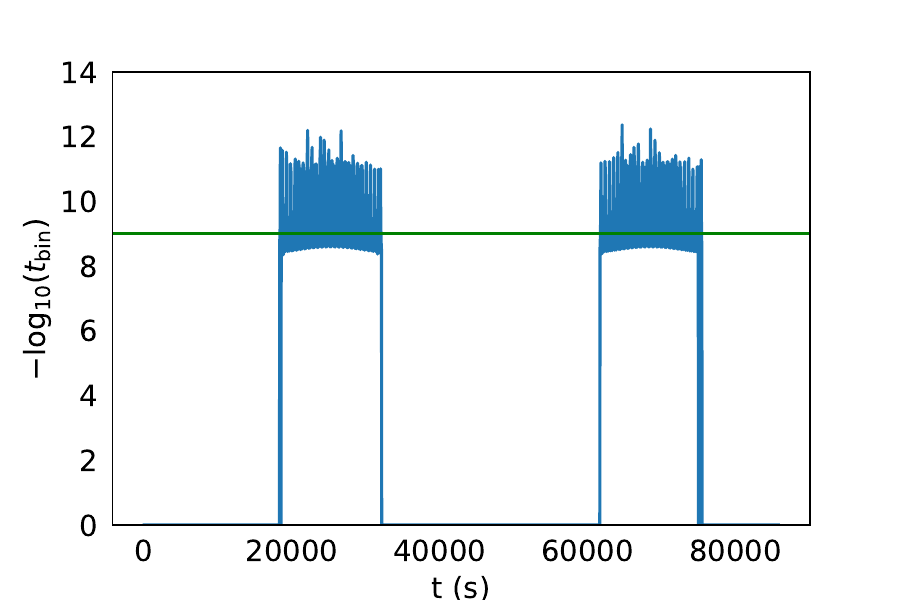}

    \caption{Sync precision versus time for various city pairs with 5 minutes of holdover time. The simulation spans one day. The green (horizontal) line represents the 1 ns cut-off, introduced by the satellite clock's precision within the holdover time. Figures on the left represent the sync precision when the satellites in an orbit are unsynchronized, whereas on the right we consider a constellation with intra-orbit sync capabilities. Clearly New York City-Montreal(top) pair retains the same precision regardless of intra-orbit connectivity as explained above. New York City-Port Au Prince (center) significantly increases in average precision as well as portion of the day where 1ns sync is possible. New York City-Puerto Montt (bottom) can sync only when the satellites are sync with each other as they will never lie in the same shadow segment.}
    \label{fig:const-long-5minprec}
\end{figure*}

We will now present some concrete examples of QCS using city pairs along similar longitudes, and compare the sync capabilities with and without intra-orbit connectivity. We choose cities along the same longitude because adding satellites to the same orbit only increases the coverage area of our QCS network along the direction of the satellites' motion (latitude). 
As above, we simulate an orbit of 10 equally-spaced satellites in a polar orbit along the Prime Meridian, and choose other parameters as in Table  \ref{tab:table1}. 
 The precision is recorded over a period of one day, with a simulation time step of $1 \rm s$. The following ``figures of merit" are evaluated: average precision of sync events at precision better than 1ns, the largest time gap between sync events with precision better than 1ns, and the percentage of the day that ground stations are synchronize with precision better than 1ns. For cases in which either or both of the cities are not visible to any satellite, we use the term ``invis." in the data. Similarly, when sync events are possible but are $\textit{not}$ better than 1ns in precision, we report the average precision above 10 $\mu$s and neglect the remaining figures. Table \ref{const-long-fom} contains these figures of merit for sync events between New York/Montreal (New York City-Montreal), New York/Port Au Prince (New York City-Port Au Prince), and New York/Puerto Montt (New York City-Puerto Montt), for holdover times of 0, 2, 5, and 10 minutes. Fig.~\ref{fig:const-long-5minprec} shows precisions over one day for holdover time $\tau$ = 5 min for the same city pairs. Note that the recorded and plotted precision values are now reported as $-\rm log_{10}(t_{bin})$, so an increase in this value indicates an ability to sync clocks with better precision. For example, for nanosecond and microsecond precision  this quantity is equal to 9 and 6, respectively. 

\begin{table*}[ht]
\begin{tabular}{|c|c|c|c|c||c|c|c|c|}
    \hline
    \multicolumn{9}{|c|}{New York City-Montreal Figures of Merit}\\
    \hline
    \multicolumn{1}{|c|}{}&
    \multicolumn{4}{|c||}{Disconnected}&
    \multicolumn{4}{|c|}{Connected}\\
    \hline
     \textbf{Holdover} & \textbf{0 min} & \textbf{2 min} & \textbf{5 min} & \textbf{10 min} & \textbf{0 min} & \textbf{2 min} & \textbf{5 min} & \textbf{10 min}\\
     \hline
     \textbf{Average Precision}&invis&9.99&10.67&11.07&invis&9.99&10.67&11.17\\
     
     \textbf{Largest Gap (min)}& invis&494&491&486&invis&494&491&486\\
     
     \textbf{Connected Fraction (\%)}&invis&8&18&33&invis&8&18&33\\
     \hline
\end{tabular}
\\
\begin{tabular}{|c|c|c|c|c||c|c|c|c|}
    \hline
    \multicolumn{9}{|c|}{New York City-Port Au Prince Figures of Merit}\\
    \hline
    \multicolumn{1}{|c|}{}&
    \multicolumn{4}{|c||}{Disconnected}&
    \multicolumn{4}{|c|}{Connected}\\
    \hline
     \textbf{Holdover} & \textbf{0 min} & \textbf{2 min} & \textbf{5 min} & \textbf{10 min} & \textbf{0 min} & \textbf{2 min} & \textbf{5 min} & \textbf{10 min}\\
     \hline
     \textbf{Average Precision}&invis&8.30&9.13&10.72&invis&9.02&10.04&11.18\\
     
     \textbf{Largest Gap (min)}& invis&n/a&551&527&invis&560&538&527\\
     
     \textbf{Connected Fraction (\%)}&invis&n/a&2&16&invis&0.20&10&26\\
     \hline
\end{tabular}
\\
\begin{tabular}{|c|c|c|c|c||c|c|c|c|}
    \hline
    \multicolumn{9}{|c|}{New York City-Puerto Montt Figures of Merit}\\
    \hline
    \multicolumn{1}{|c|}{}&
    \multicolumn{4}{|c||}{Disconnected}&
    \multicolumn{4}{|c|}{Connected}\\
    \hline
     \textbf{Holdover} & \textbf{0 min} & \textbf{2 min} & \textbf{5 min} & \textbf{10 min} & \textbf{0 min} & \textbf{2 min} & \textbf{5 min} & \textbf{10 min}\\
     \hline
     \textbf{Average Precision}&invis&invis&invis&invis&invis&9.04&10.02&11.19\\
     
     \textbf{Largest Gap (min)}& invis&invis&invis&invis&invis&577&495&490\\
     
     \textbf{Connected Fraction (\%)}&invis&invis&invis&invis&invis&2&13&32\\
     \hline
\end{tabular}
\caption{Figures of merit for the clock synchronization of city pairs along similar longitudes for various holdover times. The precision numbers are represented as absolute values in the log scale (higher is better). We use ``invis." to mean that one or both cities is invisible to any satellite in the orbit, thus synchronization is not possible. When the cities are visible but the average achievable sync precision is below the threshold of 1ns, we change the threshold to 10 $\mu$s and do not report the largest gap or connected fraction. This is done to indicate that, while the cities are  visible to a common satellite, they are unable to achieve a precision greater than 1ns. The three regimes of intra-orbit connectivity advantages are clear from these results: New York City-Montreal does not significantly benefit from connectivity as the cities always lie in the shadow of a single satellite; New York City-Port Au Prince has some increase in all figures; the most significant benefit occurs with New York City-Puerto Montt as neither city is visible to the  shadow of a single satellite simultaneously, thus intra-orbit connection results in an exponential benefit in average sync precision.}
\label{const-long-fom}
\end{table*}

These results show that in regimes where pairs of ground stations fall outside of the same shadow, as in Fig.~\ref{fig:const-long-5minprec}(e), or on the edges of the same shadow, as in Fig.~\ref{fig:const-long-5minprec}(c), equipping the orbit with satellite to satellite sync capability greatly improves the achievable precision and network scale. On the other hand, pairs of cities which are close enough to fall in the  shadow of a single satellite  do  to get any extra  benefit from this new functionality, as expected. These results  show that benefits that could be achieved by simply equipping satellites with better clocks can also be achieved, and even surpassed, by allowing satellites to sync amongst themselves. 

\section{Multi-orbit master clock}
\label{sec:5}
In this section, we introduce another level of functionality to the satellite  constellation, namely synchronization between satellites in different orbits---
inter-orbit synchronization. This is motivated by the requirement to increase the scale of the network longitude-wise, since adding holdover and allowing intra-orbit sync only allows the extension of the network scale along the satellite's direction of motion, which for polar orbits is approximately along fixed longitudes. Thus, to provide global coverage, inter-orbit synchronization is a necessary feature.
At the same time, adding this functionality is  more challenging  than intra-orbit connectivity, because satellites in different orbits move at high velocities relative to each other, unlike ones in the same orbit which are at relative rest.
Therefore, the high precision sync achieved between satellites in the same orbit is only possible for nearby orbits. We will explore and quantify this difficulty, as well as it advantages. 

\begin{figure}[h]
    \centering
        \includegraphics[width=\linewidth]{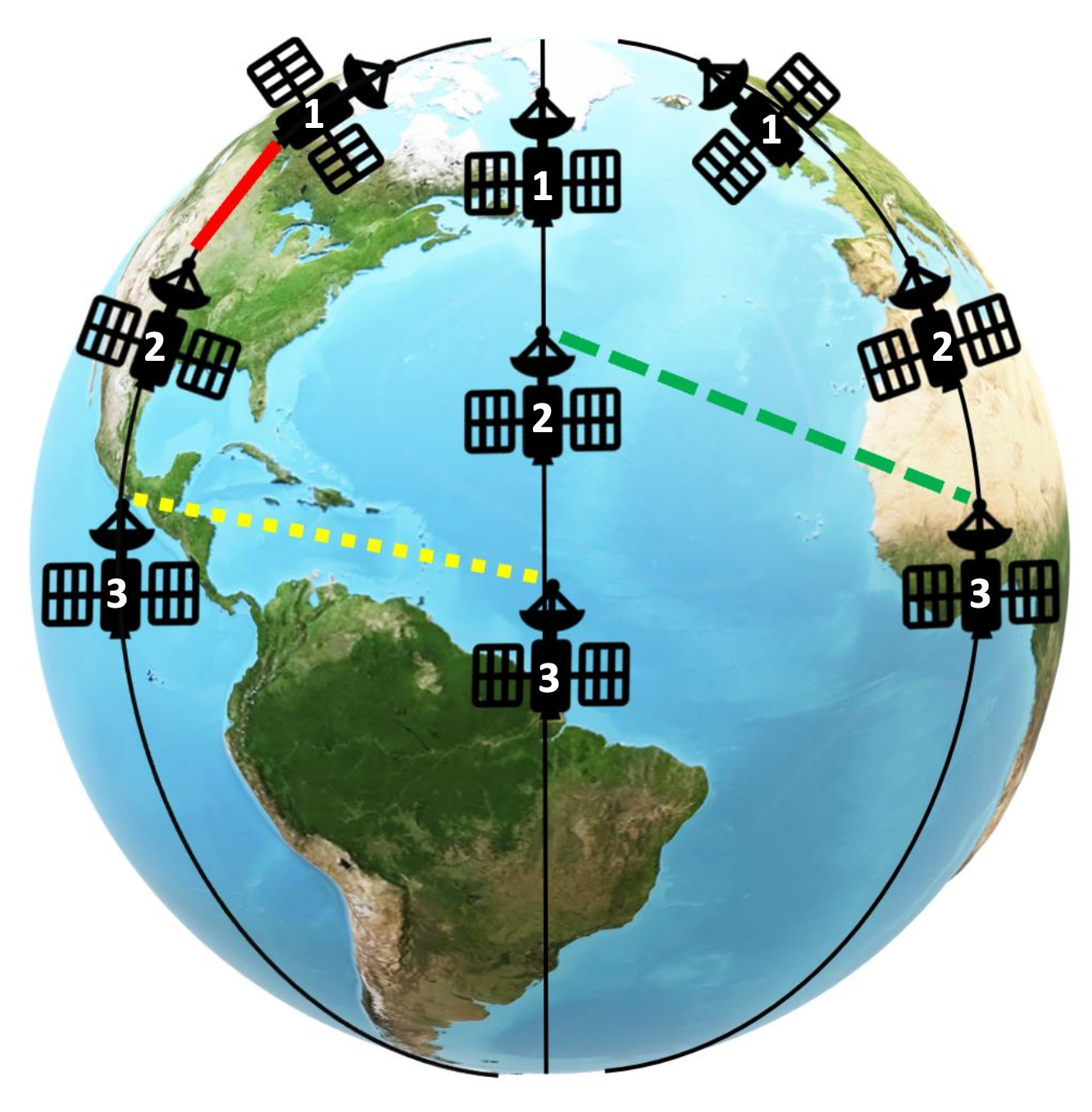}
    \caption{Possible links between satellites. The solid red line represents an intra-orbit link between satellites 1 and 2 in the same orbit, allowing a given orbit to maintain a constant sync precision between all members. The yellow dashed line is an inter-orbit link between a conjugate satellite pair (i.e. two satellites with the same number in a neighboring orbit) and the green dashed line is an inter-orbit link between two non-conjugate satellites. The conjugate pairs are the best candidates for inter-orbit sync due to their closer proximity (and thus less loss) than other pairs. Note that the simulations introduce an orbital offset of 0.25$^{\circ}$ (too small to be shown in this figure) to prevent satellite collision at the poles.}
    \label{conj_links}
\end{figure}

\subsection{Inter-orbit synchronization}
Let us begin by describing the relative dynamics of satellites in neighbouring orbits, since, analogous to the intra-orbit case, if nearest-neighbour orbits are synced all orbits are synced by transitivity. Let us consider two polar orbits both at the same altitude. Since such orbits intersect at the poles, a small offset must be added in the initial polar angles of the satellites to avoid catastrophe. Next we must ensure visibility between the satellites. This is again analogous to the single orbit case. Since 9 satellites were needed to ensure visibility amongst satellites in a ring, translating to a polar angle separation of $40^\circ$ along the orbit, the two orbits must also be no more than $40^\circ$ apart azimuthally for satellites in different orbits to see each other near the Equator. Since the Equator is the point of maximum separation between the orbits, this condition ensures visibility throughout the satellites' journey in their respective orbits.
Next, we come to the question of sync precision. We know from Eqn.~\eqref{eqn:max_precision}, that high precision can be achieved if the two clocks are very close ($\eta$, the channel transmissivity, is high) and if they move  slowly relative to each other. Intuitively, the pairs of satellites that are conjugate to each other, i.e., say the pair of satellite labeled with the 3 in each orbit (see Fig.\ref{conj_links} ), are the ones that remain closest neighbours throughout the orbital motion. Two points in the orbit are of special interest: (1) near the Poles, where the distances are minimum, but relative radial velocity is maximum, and (2) near the Equator, where conjugate satellites are at maximum separation but their velocities become almost parallel and perpendicular to the line joining them, barring the effect of the small offset. Thus, conjugate pairs of satellites are good candidates for ensuring sync between orbits. Nonetheless, this trade-off between separation and relative radial velocity makes their sync dynamics non-trivial. 

Further, if any conjugate pair between orbits can sync, the entire orbits are synced since we have already defined the conditions needed for intra-orbit sync and determined that it is continuously available at a  high level of precision. Thus, the sync precision of the master clock is limited by the inter-orbit sync capabilities, which we shall discuss now with a concrete example. 

Let us now imagine a QCS constellation with 5 orbits (this ensures constant visibility between nearest-neighbour orbits via conjugate satellites) and 10 satellites in each orbit (this ensures nearest-neighbour intra-orbit visibility). All orbits are polar and at a constant altitude of 500 km. The polar angle offset between satellites in neighbouring orbits is chosen to be $0.25^\circ$. Our aim is to determine the precision of this master clock as a function of time. 

\begin{figure}[ht]
    \centering
        \includegraphics[width=0.9\linewidth]{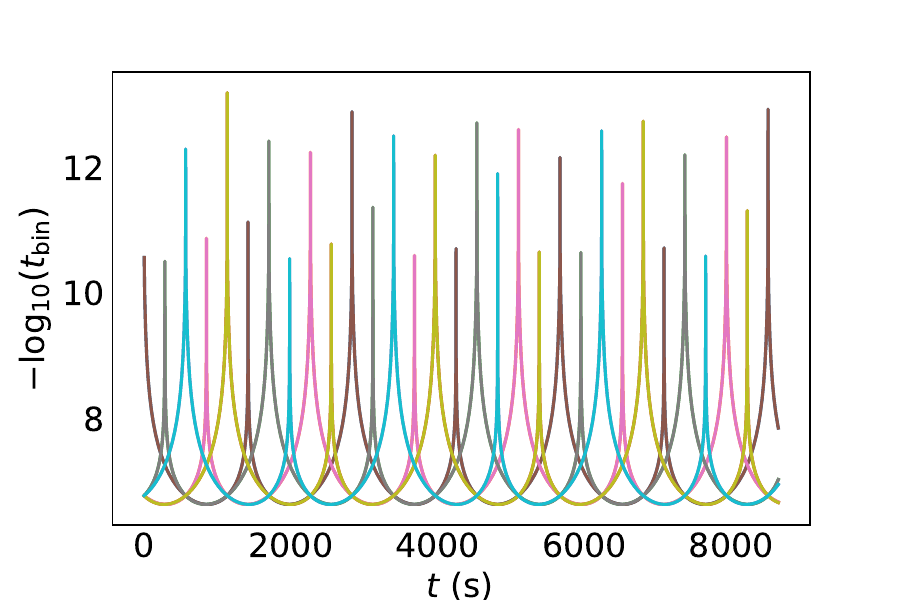}    
    \caption{Sync precision traces between orbit numbers 1 and 2 in a QCS constellation. The constellation consists of 10 satellites in 5 polar orbits each at 500 km altitude. The different colors represent precision between different conjugate pairs of satellites (E.g., precision between the pair of satellites labelled as 1 of the orbits 1 and 2 is the brown colored curve, the pair of satellites labelled as 2 is pink, etc.) All the peaks in the plot provide sub-nanosecond precision between orbit numbers 1 and 2. The gaps between these peaks are roughly 4-5 minutes, thus, a holdover time of 5 minutes for satellite clocks is enough to maintain sub-nanosecond precision continuously.}
    \label{fig:master_clock_orbit12}
\end{figure}

To this end, let us first look at the sync between conjugate satellites for a fixed pair of next-to-each-other  orbits, say orbit 1 and 2. We see in Fig.~\ref{fig:master_clock_orbit12} that high precision is obtained roughly once in 4.5 minutes. This can be understood in the following way. The members of each conjugate pair can sync with high precision four times in its orbit -- twice at the Poles and twice at the Equator. The period of the orbit is roughly 90 min, and there are 10 satellites per orbit. This gives us a gap of around 2.25 minutes between peaks. But, the behaviour of the conjugate pairs providing service to the Eastern Hemisphere is identical to the ones on the other side, and thus peaks in sync traces arising from satellites numbered 1-5 overlap with those arising from satellites numbered 6-10. Therefore, we see roughly a 4-5 minutes gap between successive peaks. Therefore, if the satellite clocks had a holdover of roughly 5 minutes the satellites in the constellation be in continuous sync at sub-nanosecond level precision.

\begin{figure}[ht]
    \centering
    
        \includegraphics[width=0.9\linewidth]{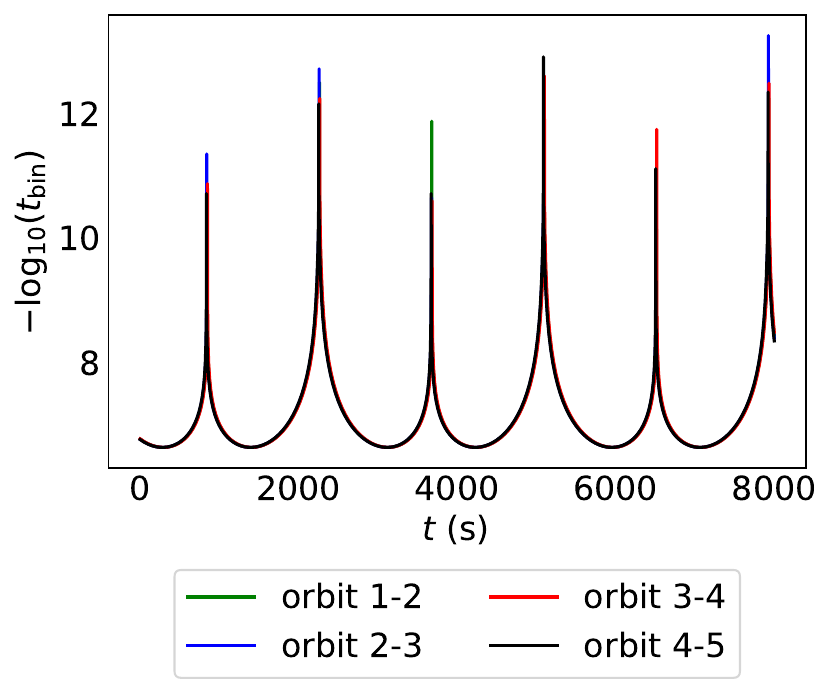}
        \includegraphics[width=0.9\linewidth]{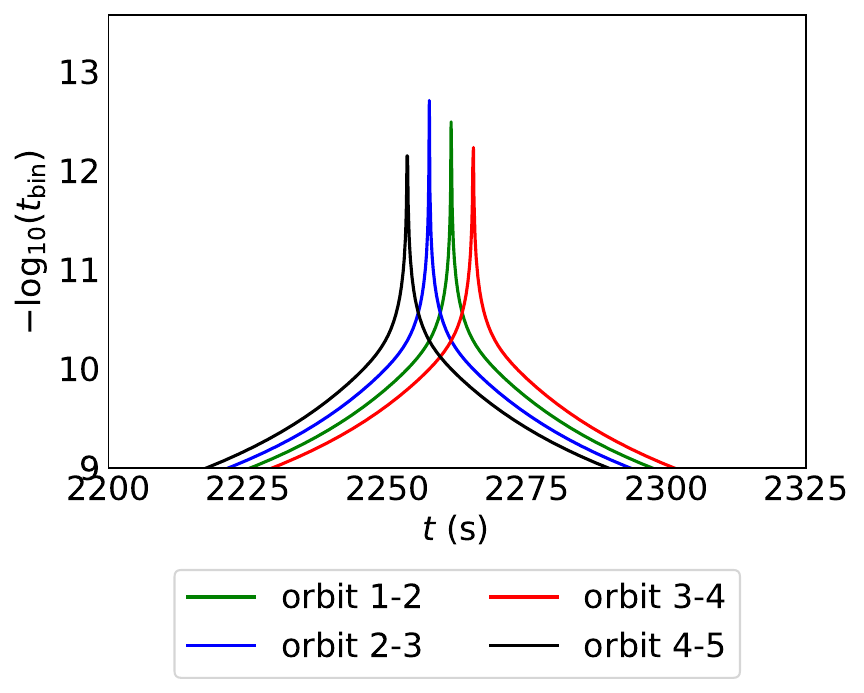}
        
    \caption{Effect of polar angle offset between conjugate satellites in neighbouring orbits on the sync precision. (Top) All orbit pairs sync at high precision and apparently also at the same time. B ut when we zoom into the peaks (Bottom) the effect of the polar offset becomes clear. For a small offset of $0.25^\circ$, the peaks for precision between different orbit pairs are displaced w.r.t. each other by not more than 30 seconds. Hence a holdover of 5 minutes required for covering the gaps between sync of different conjugate pairs is enough to also compensate this effect.}
    \label{fig:master_clock_all_orbits}
\end{figure}

 On the other hand, there is also the effect of offset between orbits. Let us look at the precision between the 1st satellite of orbit pairs 1-2, 2-3, 3-4 and 4-5. From Fig.~\ref{fig:master_clock_all_orbits} (top) it seems that all satellite pairs achieve high precision at the same time. But zooming into one of these peaks, in Fig.~\ref{fig:master_clock_all_orbits} (bottom) we find that these peaks are all slightly shifted w.r.t. each other due to the initial offset. Thus, again, to achieve high precision between different orbits, we need some holdover for the satellite clocks. In this case, a holdover of around 30 seconds is sufficient, which is much smaller than the holdover requirement discussed above and shown in Fig. (\ref{fig:master_clock_orbit12}).

In summary, we conclude that a master clock with sync precision at least at the sub-nanosecond level can be built with  constellation of 50 LEO satellites equipped with modest clocks with a small holdover time of $\approx 5$ min. Here, we must re-stress that the working principle of the master clock is to reinforce low-stability clocks forming a network through quantum links. For example, with the constellation just mentioned, clocks that can hold time at sub-nanosecond level for only a few minutes are combined to build a master clock that holds a common time indefinitely. As a consequence, the coverage area ---network scale--- on the ground also extends many-fold. Such a master clock is a primitive to a quantum network and imbibes many of the latter's essential features.

\subsection{Sync Examples}

To illustrate the enhancement in sync outcomes provided by inter-orbit sync  functionality, we present an example with three city pairs along similar latitudes, namely New York-Salt Lake City, New York-Madrid, and New York-Beijing. The constellation  is the same as discussed in the previous subsection, and we assume a modest  holdover time of 10 minutes. This choice is also motivated by the fact that it leads to a continuous 1 ns precision shadow for an orbit (see Fig.~\ref{fig:shadow_conn} (bottom)).

\begin{figure*}[ht]
    \centering
        \includegraphics[width=0.45\linewidth]
        {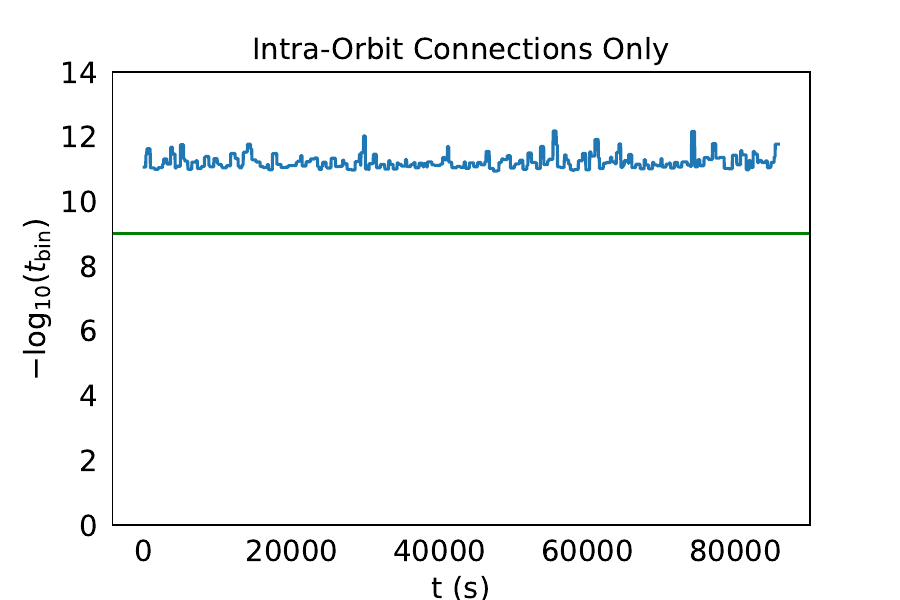}
        \includegraphics[width=0.45\linewidth]{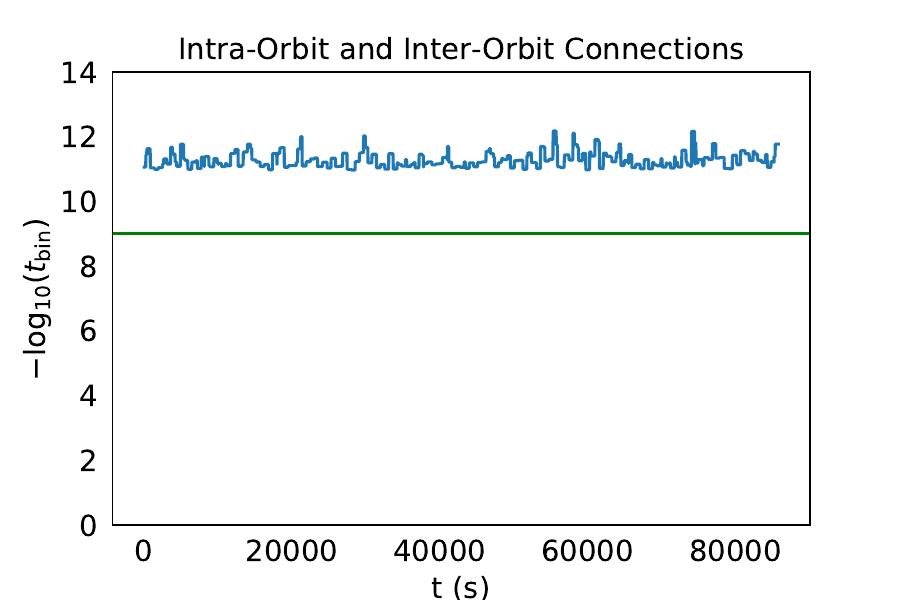}

        \includegraphics[width=0.45\linewidth]{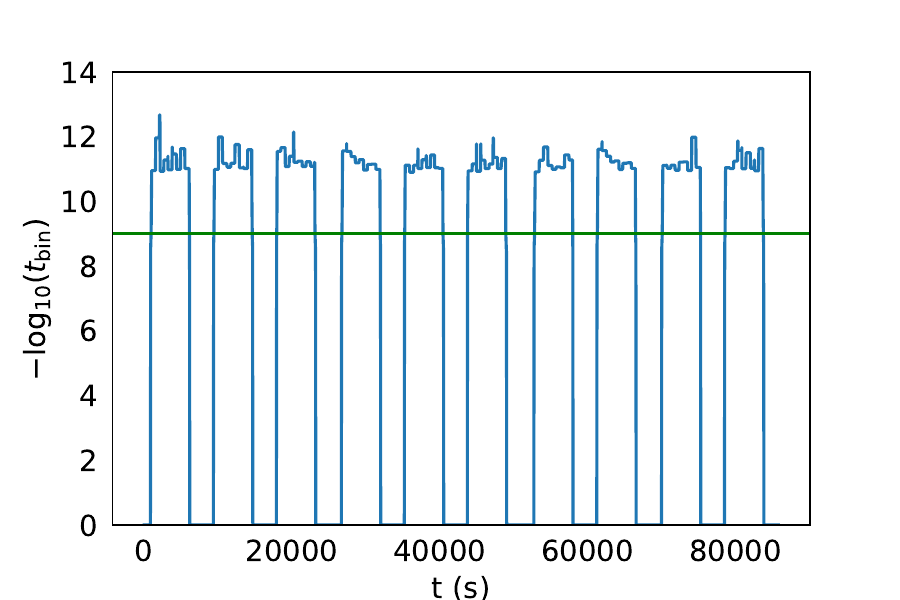}
        \includegraphics[width=0.45\linewidth]{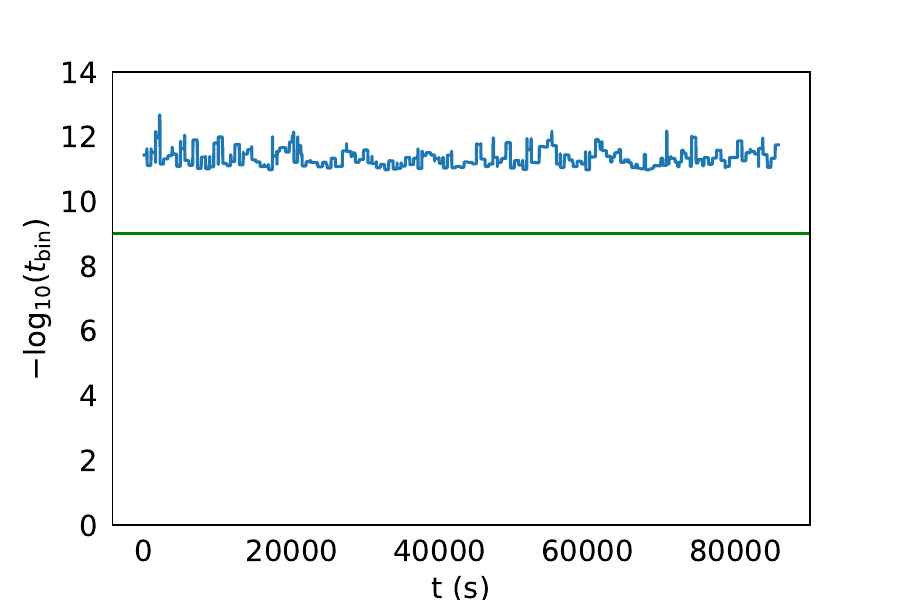}    
        
        \includegraphics[width=0.45\linewidth]{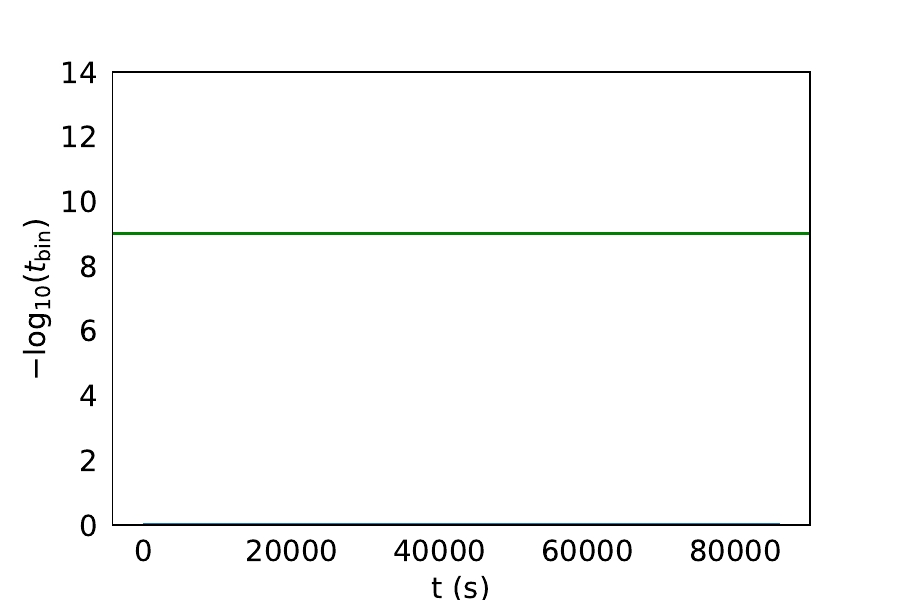}
        \includegraphics[width=0.45\linewidth]{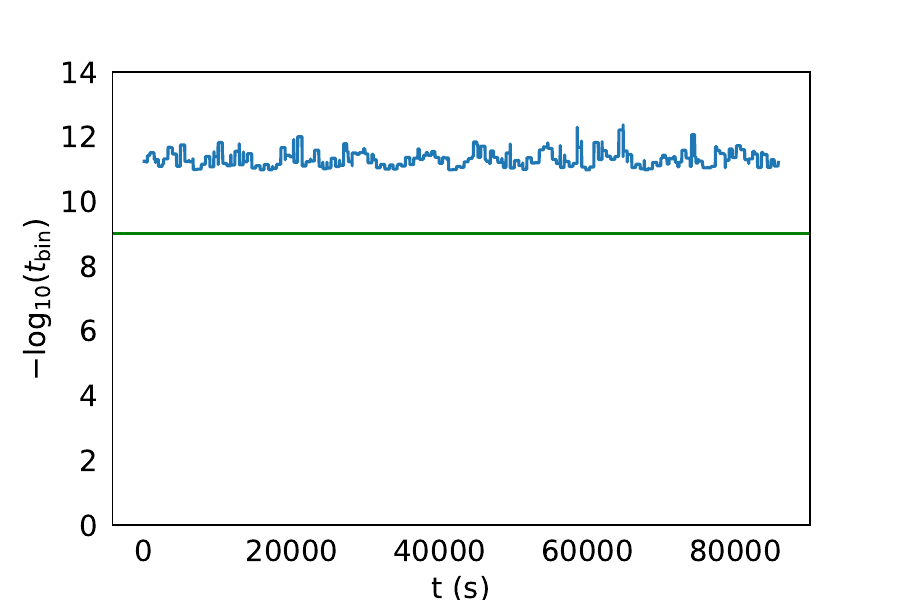}
   
    \caption{Comparison of sync precision values without (left) and with (right) inter-orbit synchronization for city pairs along similar latitudes for a constellation made of  5 orbits of 10 satellites each with $\tau$= 10 minutes. The pairs are New York-Beijing (top), New York-Salt Lake City (middle), and New York-Madrid (bottom). The horizontal green (solid) line represents the 1 ns cut-off, introduced by the satellite clock's precision within the holdover time. The plots show that, synchronization at 1 ns precision is continuously available between the selected cities. A modest improvement is seen in the sync connectivity between New York and Salt Lake City, and a drastic improvement between New York and Madrid.}
    \label{fig:const-latitude}
\end{figure*}

We are now interested in the fraction of the day for which sub-nanosecond sync is possible. Similar to the examples presented in the single-orbit case, we choose city pairs representing three distinct regimes of improvement over the case where no inter-orbit sync is assumed (but intra-orbit synchronization). We refer to the former as \emph{connected} and the latter as \emph{disconnected} case in this context. In our example, although New York and Beijing have the largest separation distance of the pairs presented, they always lie within the precision shadow of the same orbit (on the opposite sides of the Earth). Thus, their ability to synchronize is not affected by whether or not the orbits can synchronize amongst themselves. This is evident in Table \ref{tab:table2}, as both the disconnected and connected cases have 100$\%$ daily sync coverage at the 0.1 ns level (see Fig.~\ref{fig:const-latitude} (top)). This example also provides an interesting use case for a single orbit master clock. 

The next case showing an intermediate benefit from inter-orbit sync is seen through the New York-Salt Lake City pair. The cities are close enough to be visible to the same orbit but achieve sub-nanosecond sync precision only about $60\%$ of the day without inter-orbit connectivity. We see this visually in Fig. \ref{fig:const-latitude} (middle) as the precision peaks are discontinuous in the disconnected case (left) and become fully continuous once the orbits can sync (right). So by allowing inter-orbit sync, we gain an extra $40\%$ of the day in which sub-nanosecond sync is possible.
Finally, a drastic improvement  is achieved for the the New York-Madrid case. It is not possible to sync these cities without inter-orbit synchronization, as the two cities are \textit{never} simultaneously visible to the same orbit. However, since either city is always visible to \textit{some} orbit, when the orbits are continuously synced, sub-nanosecond sync between these cities can be achieved throughout the day (see Fig. \ref{fig:const-latitude} (bottom)).

\begin{table}[ht]
\begin{resizeenv}
    \begin{tabular}{|c|c|c||c|c||c|c|}
        \hline
        \multicolumn{1}{|c|}{}&
        \multicolumn{2}{|c||}{New York-Salt Lake}&
        \multicolumn{2}{|c||}{New York-Madrid}&
        \multicolumn{2}{|c|}{New York-Beijing}\\
        \hline
        \multicolumn{1}{|c|}{}&
        \multicolumn{1}{|c|}{Disconn.}&
        \multicolumn{1}{|c||}{Conn.}&
        \multicolumn{1}{|c|}{Disconn.}&
        \multicolumn{1}{|c||}{Conn.}&
        \multicolumn{1}{|c|}{Disconn.}&
        \multicolumn{1}{|c|}{Conn}
        \\
        \hline
         
         \textbf{Largest Gap (min)}&65 & 0&invis & 0& 0 &0  \\
         
         \textbf{Conn. Frac. (\%)}& 60& 100& invis& 100&100 & 100\\
         \hline
    \end{tabular}
    \end{resizeenv}
\caption{Figures of merit for the clock synchronization of city pairs along similar latitudes for the ``master clock" configuration.}
\label{tab:table2}
\end{table}

\section{Global Network}
\label{sec:6}

Now that we have discussed the advantages of  inter-satellite synchronization  and proposed the design of a master clock constellation, it remains to be shown that this clock can indeed support sub-nanosecond sync precision between any two cities on Earth. Rather than carefully selected cities along constant latitudes/longitudes as previously done, we now calculate the sync precision for pairs of cities randomly spread across 6 continents: Seattle, USA; London, United Kingdom; Rio Grande, South America; New Delhi, India; Cape Town, South Africa; and Sydney, Australia. The constellation is still the same previously discussed. 

\begin{figure}[ht]
    \centering
    \includegraphics[width=\linewidth]{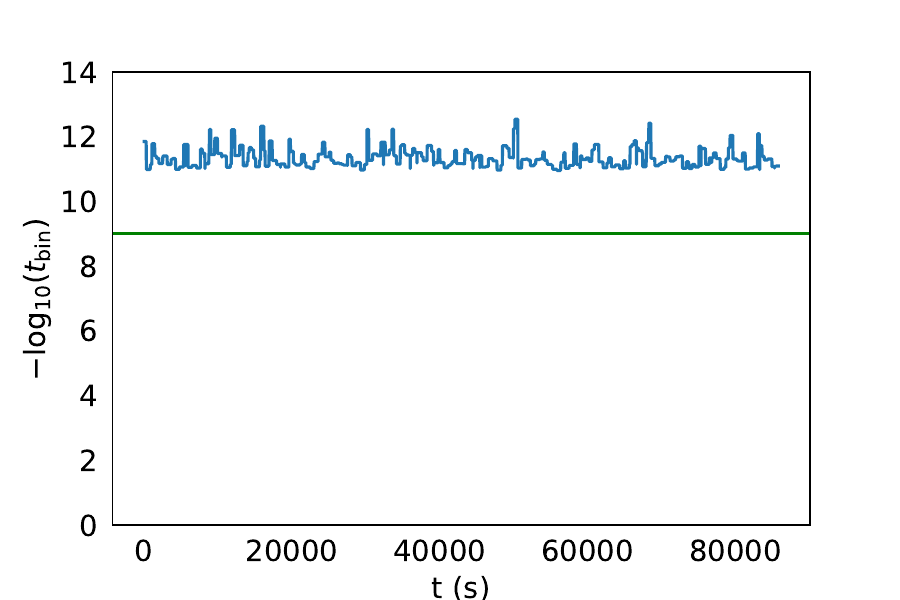}
    \includegraphics[width=\linewidth]{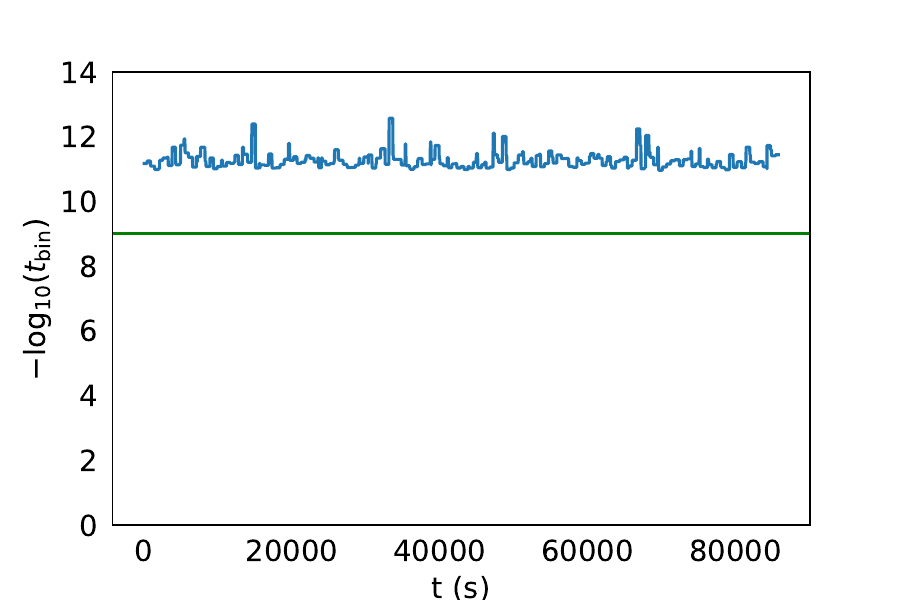}
    \includegraphics[width=\linewidth]{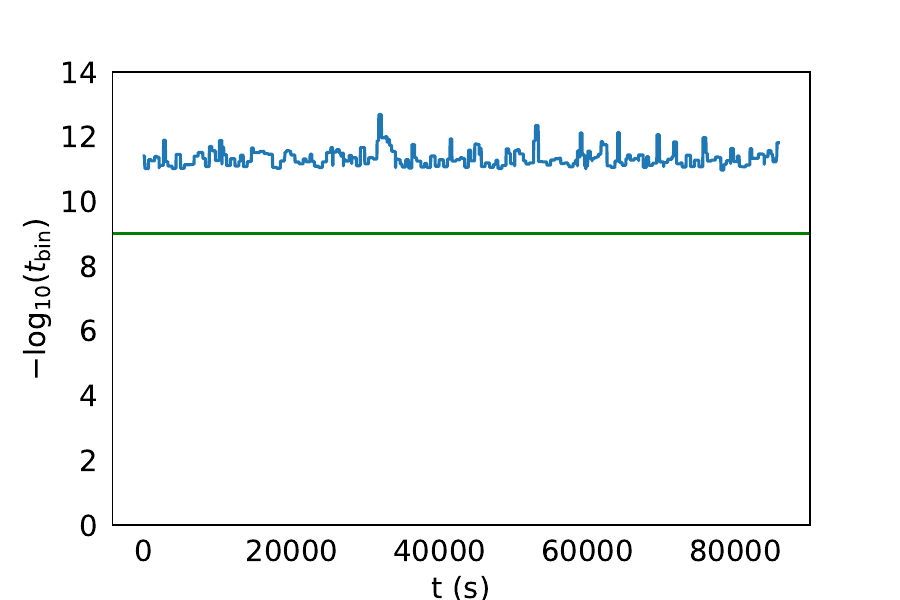}
    \caption{Sync precision over one day for a global network: Seattle-Cape Town (top;16,415 km separation), Sydney-New Delhi (middle; 10,422 km), Rio Grande-London (bottom; 7,801 km). This example provides a further proof of concept that given the previously defined master clock configuration, sub-nanosecond sync precision is achievable 100\% of the day, regardless of city separation distance.}
    \label{fig:global}
\end{figure}

Our simulations show that sub-nanosecond sync can be achieved for 100\% of the day for all city pairs for such a configuration, thus providing a continuously connected global sync network. Fig.~\ref{fig:global} affirms this for 3 city pairs, where $< 1$ ns sync precision is achievable over the entire day. However, as previously discussed, the constellation was carefully selected to have a limiting precision of 1 ns. Thus, we can conservatively conclude that a global timing network at 1 ns sync precision and continuous coverage is achievable. 

Nevertheless, these sync traces show that $10^{-10} - 10^{-11} \rm s$ level sync precision could be achieved while maintaining continuous global connectivity, if the master-clock induced nanosecond cut-off could be loosened. One way to do this is by adding more satellites to each orbit, which will reduce the holdover requirement for global coverage and also for inter-orbit QCS (see Fig. \ref{fig:master_clock_orbit12}).

\section{Conclusions}
\label{sec:7}
Enhancing the precision of time synchronization not only pushes the boundaries of fundamental science but also has important technological use cases \cite{white_paper}, both in the classical and quantum realms. Classical synchronization techniques such as those using optical frequency combs have been demonstrated to obtain femtosecond level precision, but they currently lack scalability \cite{Newbury2016_freq_combs}. The utilization of quantum resources presents an alternative avenue, accompanied by complementary advantages, and is thus deserving of exploration.
For instance, QCS comes with an added layer of quantum security which is not available in any classical protocol. Thus, in this paper, by proposing a method to extend QCS to a global scale, we provide a forward looking solution that would form the basis of future secure communications, navigation and ranging. We show that our scheme provides continuous global synchronization at the sub-nanosecond level using a quantum-assisted master clock in the sky. This is done by envisioning synchronization between satellites in the constellation. To this end, we consider inter-satellite links, show their feasibility, and quantitatively prove their ability to provide continuous and global time synchronization coverage. 

Our analyses has been done with some reasonable simplifying assumptions, as mentioned in Section \ref{sec:Intro}. We are aware that, in a real experiment, the prevalent levels of noise and jitter of channels, detectors, and time-stampers may be higher than our estimates. Current free-space QCS demonstrations require $\approx 1 \;\rm s$ of acquisition time \cite{Lanning2022, spiess_exp}, which is much more than the $\approx 1 \;\rm ms$ we use in this work, and which helps bypass the effects of relative velocities. Nonetheless, our analyses provides useful trends both qualitative and quantitative that will inform design parameters for an implementation. Although we underestimate noise, we also use very conservative parameters, representing off-the-shelf SPDC sources and detectors. Entanglement sources with rates of $\approx 10^9 - 10^{10} \; s^{-1}$ are now available and we anticipate that they would soon be space ready \cite{GHz_SPDC_sat1, GHz_SPDC_sat2}. Similarly, SNSPDs (superconducting nanowire single-photon detectors) can provide efficiencies much beyond 90\%, far outperforming simple semiconductor based detectors we have assumed \cite{1204.5560}. These improvements will allow the acquisition time to stay within the millisecond range, even with increased channel noise/loss, justifying our results in this work. Finally, here we have only focused on the use of bipartite entanglement for clock synchronization. Generation of multipartite entanglement has seen considerable technological progress recently \cite{multi_entanglement1, multi_entanglement2, multi_entanglement3, multi_entanglement4} and it would be useful to explore further how such states could be used as a tool to synchronise more than two parties at the same time (also see \cite{QClock_network_sat}). 

The main conclusions of our analysis can be summarised as follows:
\begin{itemize}
    \item QCS links between satellites in the same orbit   increase the latitudinal extent of the network. Using this minimal functionality, a high precision, single orbit master clock can be set up, since there are no relative velocities between satellites in the same orbit. Therefore, this is a functionality that can be achieved without the requirement of any holdover for the satellite clocks. Thus, surprisingly, very low stability clocks can reinforce each other within the same orbit, and then provide high precision synchronization on the ground.

    \item By allowing different orbits to synchronize amongst themselves, the extent of the network increases longitudinally. It should be noted that all other constellation enhancements, such as increasing satellite clock holdover, number of satellites, etc., could only increase the network scale along the direction of the satellite's motion. Hence, to achieve global coverage, inter-orbit synchronization is essential.

    \item Finally, as an example we have shown that a master clock comprising of 50 satellites distributed uniformly in five 500 km polar orbits, and equipped with both intra and inter orbit QCS links is able to provide sub-nanosecond precision globally, uninterrupted. This ability is beyond what is currently achievable via the GPS. 
\end{itemize}

In conclusion, the above analysis makes the proposed QCS constellation a promising precursor to a quantum version of GPS. Our proposal can also be considered as a ``quantum-assisted network of clocks". Individual nodes have low timing stability, but together they form an highly precise master clock network.

\begin{acknowledgements}
We have benefited from discussions  with Rachel McDonald. I.A., S.D. and S.H., are supported by the NSF grants PHY-2110273, by the RCS program of  Louisiana Boards of Regents through the grant LEQSF(2023-25)-RD-A-04, by funds from Xairos Inc., and by the Hearne Institute for Theoretical Physics. 
\end{acknowledgements}

\bibliography{main.bib}
\end{document}